\documentclass[letterpaper,english,reprint, aps]{revtex4-2}
\usepackage[T1]{fontenc}
\usepackage[latin9]{inputenc}
\usepackage{bm}
\usepackage{amsmath}
\usepackage{amssymb}
\usepackage{graphicx}
\usepackage{xcolor}

\makeatletter

\pdfpageheight\paperheight
\pdfpagewidth\paperwidth

\makeatother

\usepackage{babel}
\begin{document}
\title{Laser induced enhanced coupling between photons and squeezed magnons
in antiferromagnets}
\author{J. C. G. Henriques$^{1}$, T. V. C. Antão$^{1}$ and N. M. R. Peres$^{1,2}$}
\affiliation{$^{1}$Department and Centre of Physics, University of Minho, Campus
of Gualtar, 4710-057, Braga, Portugal}
\affiliation{$^{2}$International Iberian Nanotechnology Laboratory (INL), Av. Mestre
Jose Veiga, 4715-330, Braga, Portugal}
\begin{abstract}
In this paper we consider a honeycomb antiferromagnet subject to an
external laser field. Obtaining a time-independent effective Hamiltonian,
we find that the external laser renormalizes the exchange interaction
between the in-plane components of the spin-operators, and induces
a synthetic Dzyaloshinskii-Moria interaction (DMI) between second
neighbors. The former allows the control of the magnon dispersion's
bandwidth and the latter breaks time-reversal symmetry inducing non-reciprocity
in momentum space. The eigen-excitations of the system correspond
to squeezed magnons whose squeezing parameters depend on the properties
of the laser. When studying how these spin excitations couple with
cavity photons, we obtain a coupling strength which can be enhanced
by an order of magnitude via careful tuning of the laser's intensity,
when compared to the case where the laser is absent. The transmission
plots through the cavity are presented, allowing the mapping of the
magnons' dispersion relation.
\end{abstract}
\maketitle

\section{Introduction}

The field of magnon spintronics deals with the study and manipulation
of magnetic excitations, also known as magnons, in ordered magnets
\citep{chumak2015magnon,yuan2021quantum}. These spin excitations
attract considerable interest from the scientific community due to
the possibility of being used as information carriers, since they
present nanometre wavelengths and reduced losses due to Joule heating
when compared with traditional electronic devices \citep{chumak2015magnon,chumak2014magnon,demidov2011excitation,khitun2010magnonic,schneider2008realization}.
Due to the possibility of combining magnons with cavity photons \citep{soykal2010strong,harder2016study,harder2021coherent}
and superconducting qubits \citep{tabuchi2015coherent,liu2019magnon},
the study of spin excitations also found its way to the field of quantum
information \citep{bittencourt2019magnon}. The combined study of
magnon spintronics and its applications in quantum information science
gave rise to the field of quantum magnonics.

Due to their bosonic nature, magnons share various features with other
bosons, such as phonons and photons. In fact, when studying the quantum
properties of magnons, similar ideas and methods to the ones usually
found in quantum optics \citep{gerry2005introductory,fox2006quantum}
appear; one such idea is that of squeezed states. This type of quantum
state has been thoroughly studied with photons, where these states
are usually obtained out of equilibrium through four wave mixing or
parametric processes \citep{gerry2005introductory,fox2006quantum,andersen201630}.
The signature feature of squeezed states is the possibility of reducing
the uncertainty associated with a given observable by increasing the
uncertainty of another one, in such a way that, when combined, the
two uncertainties still respect Heisenberg's uncertainty principle.
This type of quantum state of light has bee used, for example, in
the detection of gravitational waves \citep{aasi2013enhanced}.

Contrarily to what is found with photons, where the study and generation
of squeezed states is a mature field \citep{walls1983squeezed}, the
study of magnon squeezing has only began to gain traction recently.
This topic, however, is a rather interesting one due to the large
values of squeezing found in these systems \citep{kamra2019antiferromagnetic},
the increased spin carried by a squeezed magnon \citep{kamra2016super},
enhanced magnon-magnon coupling \citep{liensberger2019exchange},
magnon entanglement \citep{zou2020tuning}, among others \citep{erlandsen2019enhancement}.
Moreover, contrarily to photons, the exploration of such phenomena
in magnetically ordered systems allows for its implementation in on-chip
nanodevices \citep{kamra2020magnon}.

Regarding the generation of squeezed magnons, two distinct approaches
can be employed. On the one hand, squeezing can be obtained by driving
the magnetic system out of equilibrium while coupled with an optomechanical
cavity \citep{li2019squeezed,yang2021bistability,zhang2021generation}.
On the other hand, and in stark contrast with photons, anisotropic
ferromagnets and isotropic antiferromagnets host equilibrium magnon
squeezing, that is, the eigen-excitations of such systems naturally
present squeezing, making them robust against environment perturbations
\citep{yuan2021quantum,kamra2019antiferromagnetic,kamra2020magnon}.

Recent works focus on the manipulation of magnetic materials via the
application of high-frequency laser fields. In Ref. \citep{Hirosawa2022},
the authors uncover an ultrafast Floquet magnonic topological phase
transition in a laser-driven skyrmion crystal, and demonstrate how
single skyrmions can be set in motion with a velocity and propagation
direction that can be tuned by the laser. In Ref. \citep{owerre2017floquet}
a study of Floquet topological magnons in ferromagnets was performed,
where the author explored how the application of an external laser
field may be used to generate a synthetic Dzyaloshinskii-Moria interaction
(DMI) \citep{kim2016realization} through the appearance of a time-dependent
Aharonov-Casher phase \citep{aharonov1988comment}. This laser induced
effect lead to the transformation of Dirac magnons into magnon Chern
insulators. Inspired by this work, and motivated by the growing interest
of the scientific community on the magnonic response of antiferromagnets,
in this paper we study the effect of applying an external laser field
to a honeycomb antiferromagnet, and discuss how it can be used to
tune the properties of the squeezed magnons hosted by the system.

The text is organized as follows: In Sec. \ref{sec:Model-Hamiltonian}
we start by introducing the Hamiltonian of a honeycomb antiferromagnt,
and how it is modified by the presence of the laser. Afterwards the
Hamiltonian is diagonalized with a Bogoliubov transformation. In Sec.
\ref{sec:Magnon-photon-coupling} we consider the system to be placed
in an optical cavity, and study how the laser can be used to enhance
the magnon-photon coupling; plots of the transmission through the
cavity are also given. In Sec. \ref{sec:Plausible-parameters} we
discuss the feasibility of experimentally realizing this type of system,
and in Sec. \ref{sec:Final-remarks} we give our final remarks.

\begin{figure}[h]
\centering{}\includegraphics[scale=1.1]{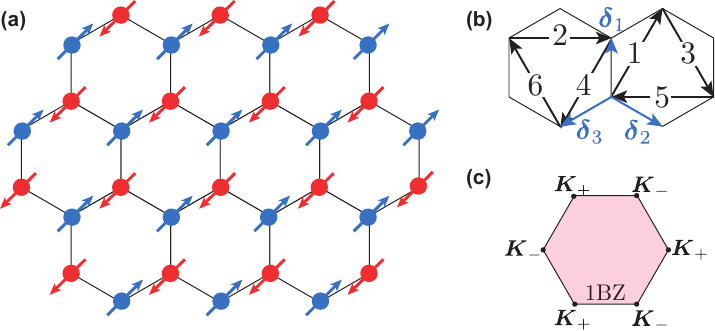}\caption{\label{fig:Lattice}(a) Schematic representation of the antiferromagnet
honeycomb lattice. The distance between nearest neighbors is $a$.
(b) Representation of the three $\boldsymbol{\delta}$ vectors connecting
one site to its three nearest neighbors. Also depicted are the 6 vectors
$\boldsymbol{b}_{i}$ which enter in the definition of the Dzyaloshinskii-Moria
interaction. (c) Schematic representation of the first Brillouin zone
(1BZ), with the Dirac points $a\boldsymbol{K}_{\pm}=\pm2\pi\left(2/3\sqrt{3},0\right)$
highlighted.}
\end{figure}

\section{Model Hamiltonian\label{sec:Model-Hamiltonian}}

In this section, we introduce the Hamiltonian of a honeycomb antiferromagnet. We focus on this type of lattice since it is a common one in two-dimensional magnetic materials, and as such is relevant in this field of research. 
Afterwards, the effect of applying a circularly polarized laser field
to the system will be explored, with the introduction of a Floquet
effective Hamiltonian. This process is analogous to the one presented
for a ferromagnet in \citep{owerre2017floquet}. Finally, the resulting
Hamiltonian will be diagonalized, and its properties discussed.

\subsection{Antiferromagnetic Synthetic Hamiltonian}

Consider the antiferromagnetic honeycomb lattice depicted in Fig.
\ref{fig:Lattice}. Accounting only for nearest neighbor interactions,
and assuming an identical easy axis anisotropy for both sublattices
($A$ and $B$), we write
\begin{align}
H & =J\sum_{i\in A}\sum_{j=1}^{3}\boldsymbol{S}_{A}(\boldsymbol{r}_{i})\cdot\boldsymbol{S}_{B}(\boldsymbol{r}_{i}+\boldsymbol{\delta}_{j})\nonumber \\
 & +K\sum_{i\in A}S_{A}^{z}(\boldsymbol{r}_{i})S_{A}^{z}(\boldsymbol{r}_{i})+K\sum_{i\in B}S_{B}^{z}(\boldsymbol{r}_{i})S_{B}^{z}(\boldsymbol{r}_{i})
\end{align}
where $J>0$ is the antiferromagnetic exchange coupling, $K$ quantifies
the easy axis anisotropy, $\boldsymbol{S}_{A/B}=(S_{A/B}^{x},S_{A/B}^{y},S_{A/B}^{z})$
are the spin operators of the sublattice $A/B$ and $\boldsymbol{\delta}_{1}=a\left(0,1\right)$,
$\boldsymbol{\delta}_{2}=a\left(1,-\sqrt{3}\right)/2$, $\boldsymbol{\delta}_{3}=-a\left(1,\sqrt{3}\right)/2$
are the vectors connecting each site of sublattice $A$ site to its
three nearest neighbors, belonging to sublattice $B$, with $a$ the
nearest neighbor distance. This Hamiltonian is commonly employed to
study a variety of antiferromagnets, and can be extended to include
effects we neglect here, such as an in-plane anisotropy.

Inspired by \citep{owerre2017floquet}, we now consider that a circularly
polarized laser field is applied to this system. The electric field
component associated with the laser is given by $\boldsymbol{E}_{0}=E_{0}\left(\tau\cos\omega_{l}t,\sin\omega_{l}t,0\right)$,
where $\tau=\pm1$ encodes the two possible circular polarizations
and $\omega_{l}$ stands for the laser's frequency. Now, following
\citep{owerre2017floquet}, we consider that the laser couples with
the system through the Aharonov-Casher effect \citep{aharonov1988comment}.
This effect, which can be viewed as complementary to the Aharonov-Bohm
effect, presents itself by giving rise to a complex phase due to the
interaction of a magnetic momentum with an electric field. For the
system at hand, this translates to the appearance of a time dependent
complex exponential, termed the Aharanov-Casher phase. This time dependent phase is a generalization of what is found in the static limit \citep{meier2003magnetization}, and has been discussed before in several works \citep{elyasi2019topologically, kar2018photoinduced, owerre2019photo, owerre2019photoinduced, owerre2019magnonic, proskurin2019microscopic, vinas2020light}. The Hamiltonian
of the system in such a situation becomes time dependent, and may
be expressed as
\begin{align}
H(t) & =J\sum_{i\in A}\sum_{j=1}^{3}S_{A}^{z}(\boldsymbol{r}_{i})S_{B}^{z}(\boldsymbol{r}_{i}+\boldsymbol{\delta}_{j})\nonumber \\
 & +\frac{J}{2}\sum_{i\in A}\sum_{j=1}^{3}\left[S_{A}^{+}(\boldsymbol{r}_{i})S_{B}^{-}(\boldsymbol{r}_{i}+\boldsymbol{\delta}_{j})e^{-i\theta_{\boldsymbol{r}_{i},\boldsymbol{r}_{i}+\boldsymbol{\delta}_{j}}}+\textrm{h.c.}\right]\nonumber \\
 & +K\sum_{i\in A}S_{A}^{z}(\boldsymbol{r}_{i})S_{A}^{z}(\boldsymbol{r}_{i})+K\sum_{i\in B}S_{B}^{z}(\boldsymbol{r}_{i})S_{B}^{z}(\boldsymbol{r}_{i})
\end{align}
where we introduced $S_{\alpha}^{\pm}=S_{\alpha}^{x}\pm S_{\alpha}^{y}$
with $\alpha=A,B$. The time dependent Aharonov-Casher phase is defined
as 
\begin{equation}
\theta_{\boldsymbol{r}_{i},\boldsymbol{r}_{i}+\boldsymbol{\delta}_{j}}=\lambda\sin\left(\omega_{l}t-\tau\phi_{\boldsymbol{r}_{i},\boldsymbol{r}_{i}+\boldsymbol{\delta}_{j}}\right),
\end{equation}
with $\lambda=\tau g\mu_{B}E_{0}a/\hbar c^{2}$ where $\mu_{B}$ is
the Bohr magneton and $g$ the Landé g-factor; $\phi_{\boldsymbol{r}_{i},\boldsymbol{r}_{i}+\boldsymbol{\delta}_{j}}$
corresponds to the angle between the vectors $\boldsymbol{r}_{i}$
and $\boldsymbol{r}_{i}+\boldsymbol{\delta}_{j}$. Note that since
$\lambda\propto E_{0}$ the value of $\lambda$ is proportional to
the laser's intensity. To avoid working with a time dependent Hamiltonian
we shall make use of Floquet theory \citep{sentef2015theory,rechtsman2013photonic,wang2013observation,lindner2011floquet}.
This is a perturbative approach allows us to obtain an effective time
independent Hamiltonian from $H(t)$. This is achieved by expressing
the effective Hamiltonian in a high frequency series expansion, which
up to first order reads
\begin{equation}
H^{\textrm{eff}}=H^{(0)}+\sum_{m=1}^{\infty}\frac{\left[H^{(m)},H^{(-m)}\right]}{m\hbar\omega_{l}},
\end{equation}
where $H^{(m)}$ is the $m$-th Fourier component of $H(t)$. In order
for this effective Hamiltonian to be valid, the energy of the laser
should be larger than the energy scale of the initial system (set
by $J$). For lower laser energies more terms would have to be accounted
for when defining $H^{\textrm{eff}}$. The leading term $H^{(0)}$
corresponds, perhaps quite unsurprisingly, to an average of the Hamiltonian
time dependent Hamiltonian over a period of the driving laser field.
It is the most relevant term for our purposes, with the first order
correction introducing only small effects.

We shall work within the linear spin-wave theory. Thus, we introduce
the linearized Holstein-Primakoff transformations, which for the antiferromagnetic
case read \citep{pires2021theoretical}
\begin{align}
S_{A}^{+}(\boldsymbol{r}_{i}) & =\sqrt{2S}a_{\boldsymbol{r}_{i}},\quad S_{A}^{z}(\boldsymbol{r}_{i})=S-a_{\boldsymbol{r}_{i}}^{\dagger}a_{\boldsymbol{r}_{i}}\\
S_{B}^{+}(\boldsymbol{r}_{i}) & =\sqrt{2S}b_{\boldsymbol{r}_{i}}^{\dagger},\quad S_{B}^{z}(\boldsymbol{r}_{i})=-S+b_{\boldsymbol{r}_{i}}^{\dagger}b_{\boldsymbol{r}_{i}},
\end{align}
where $a_{\boldsymbol{r}_{i}}/a_{\boldsymbol{r}_{i}}^{\dagger}$ and
$b_{\boldsymbol{r}_{i}}/b_{\boldsymbol{r}_{i}}^{\dagger}$ are bosonic
operators, which we refer to as the annihilation/creation operators
of sublattice magnons, and we note $\left(S_{A/B}^{\pm}\right)^{\dagger}=S_{A/B}^{\mp}$.
Introducing the Fourier representation of the Holstein-Primakoff annihilation/creation
operators, $a_{\boldsymbol{r}_{i}}=\frac{1}{\sqrt{N}}\sum_{\boldsymbol{k}}e^{i\boldsymbol{k}\cdot\bm{r}_{i}}a_{\boldsymbol{k}}$,
we obtain the following time-independent effective Hamiltonian in
momentum space
\begin{align}
H^{\textrm{eff}}= & JS\sum_{\boldsymbol{k}}\left(3+\frac{2K}{J}\right)\left(a_{\boldsymbol{k}}^{\dagger}a_{\boldsymbol{k}}+b_{\boldsymbol{k}}^{\dagger}b_{\boldsymbol{k}}\right)\nonumber \\
+ & JS\mathcal{J}_{0}\left(\lambda\right)\sum_{\boldsymbol{k}}\left(a_{\boldsymbol{k}}b_{\boldsymbol{-k}}\phi_{\boldsymbol{k}}+a_{\boldsymbol{k}}^{\dagger}b_{-\boldsymbol{k}}^{\dagger}\phi_{\boldsymbol{k}}^{*}\right)\nonumber \\
- & JS\sum_{\boldsymbol{k}}D_{\boldsymbol{k}}\left(\omega_{l},\lambda\right)b_{\boldsymbol{k}}^{\dagger}b_{\boldsymbol{k}}+JS\sum_{\boldsymbol{k}}D_{\boldsymbol{k}}\left(\omega_{l},\lambda\right)a_{\boldsymbol{k}}^{\dagger}a_{\boldsymbol{k}}\label{eq:H_eff}
\end{align}
with $\phi_{\boldsymbol{k}}=\sum_{j=1}^{3}e^{-i\boldsymbol{k}\cdot\boldsymbol{\delta}_{j}}$
a geometric factor stemming from the nearest neighbor interactions.
The first order correction arises in the previous equation in the
form of the coefficients
\begin{equation}
D_{\boldsymbol{k}}(\omega_{l},\lambda)=\frac{4JS}{\hbar\omega_{l}}\mathcal{V}_{\textrm{DM}}\left(\lambda\right)\textrm{Im}\sum_{j=1}^{6}(-1)^{j}e^{-i\boldsymbol{k}\cdot\boldsymbol{b}_{j}},
\end{equation}
where the $\boldsymbol{b}_{j}$ are shown in Fig. \ref{fig:Lattice}
and correspond to the vectors connecting a given site to its six next
nearest neighbors; $\mathcal{V}_{\textrm{DM}}\left(\lambda\right)=\sum_{m\geq1}^{\infty}\left[\mathcal{J}_{m}^{2}(\lambda)/2m\right]\sin\left(\tau m2\pi/3\right)$
and $\mathcal{J}_{m}(x)$ is the cylindrical Bessel function of the
first kind of order $m$. The terms proportional to $D_{\boldsymbol{k}}$
correspond to a synthetic Dzyaloshinskii-Moria interaction (DMI),
which appears from laser induced couplings between second neighbors;
its magnitude can be tuned simultaneously through the laser's intensity
and frequency. Thus, comparing this Hamiltonian with the one usually
employed to describe an antiferromagnet \citep{kamra2019antiferromagnetic},
we realize that: (i) the exchange coupling between the $z$-component
of the spins and the easy axis anisotropy are unaffected by the laser
and originate the first line of Eq. (\ref{eq:H_eff}); (ii) the laser
is responsible for the renormalization of the in-plane exchange coupling
between nearest neighbors, reducing its strength, i.e. $J\rightarrow J\mathcal{J}_{0}(\lambda)$ (this renormalization would also appear on other lattice configurations),
and finally (iii) a synthetic DMI arises from the laser-induced interaction
between next nearest neighbors. We note in passing that although the DMI is finite for the honeycomb lattice, that is not always the case, since, for example, it is absent for square lattices. How these modifications manifest themselves
on the properties of the system is studied below.

\subsection{Diagonalization}

To diagonalize the effective Hamiltonian $H^{\textrm{eff}}$ a Bogoliubov
transformation shall be used, as often is the case when working with
antiferromagnets. Hence, to achieve this, we introduce two new sets
of operators, $\alpha_{\boldsymbol{k}}/\alpha_{\boldsymbol{k}}^{\dagger}$
and $\beta_{\boldsymbol{k}}/\beta_{\boldsymbol{k}}^{\dagger}$, which
we simply refer to as magnon annihilation/creation operators. These
are defined as:
\begin{align}
\alpha_{\boldsymbol{k}} & =u_{\boldsymbol{k}}a_{\boldsymbol{k}}+v_{\boldsymbol{k}}b_{-\boldsymbol{k}}^{\dagger}\\
\beta_{\boldsymbol{k}} & =v_{\boldsymbol{k}}a_{\boldsymbol{k}}^{\dagger}+u_{\boldsymbol{k}}b_{-\boldsymbol{k}},
\end{align}
with $u_{\boldsymbol{k}}=\cosh\frac{\xi_{\boldsymbol{k}}}{2}$ and
$v_{\boldsymbol{k}}=e^{i\theta_{\boldsymbol{k}}}\sinh\frac{\xi_{\boldsymbol{k}}}{2}$
where $\xi_{\boldsymbol{k}}>0$ and $\theta_{\boldsymbol{k}}\in[0,2\pi)$.
Imposing that $H^{\textrm{eff}}$ is diagonal when expressed in terms
of $\alpha_{\boldsymbol{k}}$ and $\beta_{\boldsymbol{k}}$, that
is, $H^{\textrm{eff}}=\sum_{\boldsymbol{k}}\epsilon_{\alpha,\boldsymbol{k}}\alpha_{\boldsymbol{k}}^{\dagger}\alpha_{\boldsymbol{k}}+\epsilon_{\beta,\boldsymbol{k}}\beta_{\boldsymbol{k}}^{\dagger}\beta_{\boldsymbol{k}}$,
we find the following dispersion relation up to an overall constant
factor:
\begin{equation}
\epsilon_{\alpha,\boldsymbol{k}}=\epsilon_{\beta,\boldsymbol{k}}=JS\left(3+D_{\boldsymbol{k}}+\frac{2K}{J}\right)\sqrt{1-\tanh^{2}\xi_{\boldsymbol{k}}}\label{eq:magnon_disp}
\end{equation}
for
\begin{equation}
\tanh^{2}\xi_{\boldsymbol{k}}=\left(\frac{\mathcal{J}_{0}\left(\lambda\right)}{3+D_{\boldsymbol{k}}+2K/J}\right)^{2}|\phi_{\boldsymbol{k}}|^{2},
\end{equation}
and
\begin{equation}
\theta_{\boldsymbol{k}}=n\pi-\arctan\frac{\textrm{Im\ensuremath{\phi_{\boldsymbol{k}}}}}{\textrm{Re}\ensuremath{\phi_{\boldsymbol{k}}}},\label{eq:Bogoliubov_theta_k}
\end{equation}
where $n=0,1$ is fixed through the condition $\textrm{sign}\left[\cos\theta_{\boldsymbol{k}}\right]=\textrm{sign}\left[\mathcal{J}_{0}\left(\lambda\right)\right]$.

\subsubsection{Energy dispersion}

From the diagonalization procedure, we find that the two magnon modes
of the system are degenerate, i.e. $\epsilon_{\alpha,\boldsymbol{k}}=\epsilon_{\beta,\boldsymbol{k}}$.
This is known to be the case for usual antiferromagntes, and we verify
that the introduction of the laser field does not change this aspect
of the problem. This degeneracy can, however, be broken in several
manners, for example by defining the easy axis anisotropy as being
different for the two sublattices, applying a magnetic field along
the $z$-direction, or accounting for magnetic dipolar interactions
\citep{kamra2017noninteger}.

The effect of the applied laser field on the magnon dispersion is
twofold. If the laser energy is tuned such that $\hbar\omega_{l}\gg J$,
then the synthetic DMI essentially vanishes, $D_{\boldsymbol{k}}\approx0$,
and the laser manifests itself through the term $\mathcal{J}_{0}(\lambda)$
only. This term is responsible for modifying the bandwidth of the
magnon dispersion, which now oscillates with steadily decaying amplitude
as $\lambda$ increases, similarly to the behavior of a Bessel function.
Since $\phi_{\boldsymbol{k}}$ is maximal at $\boldsymbol{k}=0$,
and vanishes at the Dirac points, the effect of $\mathcal{J}_{0}(\lambda)$
is more pronounced at the center of the Brillouin zone, and does nothing
at its vertices, $\boldsymbol{K}_{\pm}$. 
\begin{figure}[h]
\centering{}\includegraphics{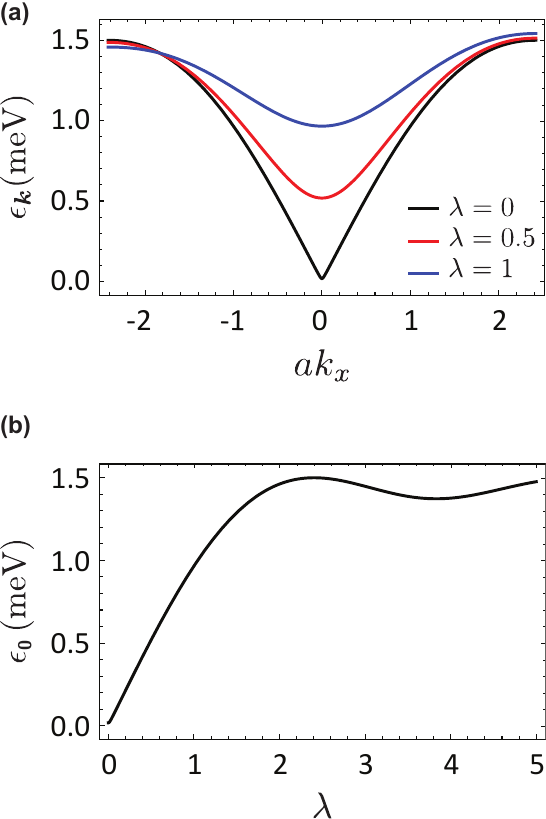}\caption{\label{fig:magnon_dispersion}(a) Energy dispersion $\epsilon_{\boldsymbol{k}}\equiv\epsilon_{\alpha,\boldsymbol{k}}=\epsilon_{\beta,\boldsymbol{k}}$
along the $k_{y}=0$ line for three different values of $\lambda$.
(b) Dependence of the $\boldsymbol{k}=0$ magnon mode energy, $\epsilon_{\boldsymbol{0}}$,
with $\lambda$. In both panels we considered $J=1$meV, $K=10^{-4}J$
and $\hbar\omega_{l}=10J$.}
\end{figure}
This behavior is illustrated in Fig. \ref{fig:magnon_dispersion}a
where we depict the magnon dispersion in momentum space along the
line $k_{y}=0$ for three distinct values of $\lambda$. In the same
figure, we also plot the magnon dispersion at $\boldsymbol{k}=0$
as a function of $\lambda$, perfectly illustrating the aforementioned
oscillating behavior. Furthermore, we also note that for a judicious
choice of $\lambda$, corresponding to the zeros of $\mathcal{J}_{0}(\lambda)$,
one obtains an almost flat band for the magnon dispersion. This is
easily understood from the inspection of Eq. (\ref{eq:magnon_disp}),
where one sees that if $\mathcal{J}_{0}(\lambda)=0$, then the square
root term simply equals one, and the momentum dependence comes only
from the term $D_{\boldsymbol{k}}$, which can be made arbitrarily
small by increasing the energy of the laser.

Consider now the case where the laser's energy is reduced, such that
the synthetic DMI becomes relevant. In this limit, time-reversal symmetry
is broken, and the dispersion relation is no longer even in momentum
space, meaning that $\epsilon_{\alpha,\boldsymbol{k}}\neq\epsilon_{\alpha,-\boldsymbol{k}}$.
Since $D_{\boldsymbol{k}=0}=0$, the effect of the DMI is mainly noticeable
near the Dirac points, where it induces an energy difference between
the magnon dispersion at $\boldsymbol{K}_{+}$ and $\boldsymbol{K}_{-}$.
This effect is visible in Fig. \ref{fig:magnon_dispersion}a, where
we observe that as $\boldsymbol{k}\rightarrow\boldsymbol{K}_{+}$
the magnon dispersion increases when compared with the case where
the laser is turned off $(\lambda=0$); the opposite statement is
valid when $\boldsymbol{k}\rightarrow\boldsymbol{K}_{-}$. The presence
of the DMI is not guaranteed to always yield this effect, since it
depends on the value of $\mathcal{V}_{\textrm{DM}}$ which may be
positive, negative, or zero, depending on the value of $\lambda$.

\subsubsection{Bogoliubov coefficients}

Now that the energy dispersion of the system was studied, let us discuss
the Bogoliubov transformation itself, and how its coefficients depend
on the laser field.

We start by noting that the definitions we gave for $\alpha_{\boldsymbol{k}}$
and $\beta_{\boldsymbol{k}}$ in terms of $a_{\boldsymbol{k}}$ and
$b_{-\boldsymbol{k}}^{\dagger}$ could be alternatively expressed
in terms of a two mode squeeze operator $S_{2}(\zeta_{\boldsymbol{k}})=\exp\left(\zeta_{\boldsymbol{k}}a_{\boldsymbol{k}}b_{-\boldsymbol{k}}-\zeta_{\boldsymbol{k}}^{*}a_{\boldsymbol{k}}^{\dagger}b_{-\boldsymbol{k}}^{\dagger}\right)$,
where $\zeta_{\boldsymbol{k}}=\left(\xi_{\boldsymbol{k}}/2\right)e^{-i\theta_{\boldsymbol{k}}}$
\citep{gerry2005introductory,fox2006quantum}. Using this operator,
we could have defined $\alpha_{\boldsymbol{k}}=S_{2}(\zeta_{\boldsymbol{k}})a_{\boldsymbol{k}}S_{2}^{\dagger}(\zeta_{\boldsymbol{k}})$
and $\beta_{\boldsymbol{k}}=S_{2}(\zeta_{\boldsymbol{k}})b_{-\boldsymbol{k}}S_{2}^{\dagger}(\zeta_{\boldsymbol{k}})$
(the proof of this statement is given in the Supplementary Information).
Hence, based on these new definitions, we see that the magnonic excitations
of the system ($\alpha_{\boldsymbol{k}}$ and $\beta_{\boldsymbol{k}}$),
which are linear combinations of sublattice magnons ($a_{\boldsymbol{k}}$
and $b_{-\boldsymbol{k}}$), correspond, in fact, to two-mode squeezed
magnons, with a squeezing parameter $\xi_{\boldsymbol{k}}$. Although
the identification of the eigenmodes of the antiferromagnet as being
squeezed magnons is independent of the external laser field, the presence
of the laser introduces new interesting features in the system due
to the possibility of tuning both $\xi_{\boldsymbol{k}}$ and $\theta_{\boldsymbol{k}}$,
as we shall see below.

Let us now study how the parameters $\xi_{\boldsymbol{k}}$ and $\theta_{\boldsymbol{k}}$
depend on the external laser field, and how their values change with
the momentum $\boldsymbol{k}$. 
\begin{figure}
\centering{}\includegraphics[scale=0.8]{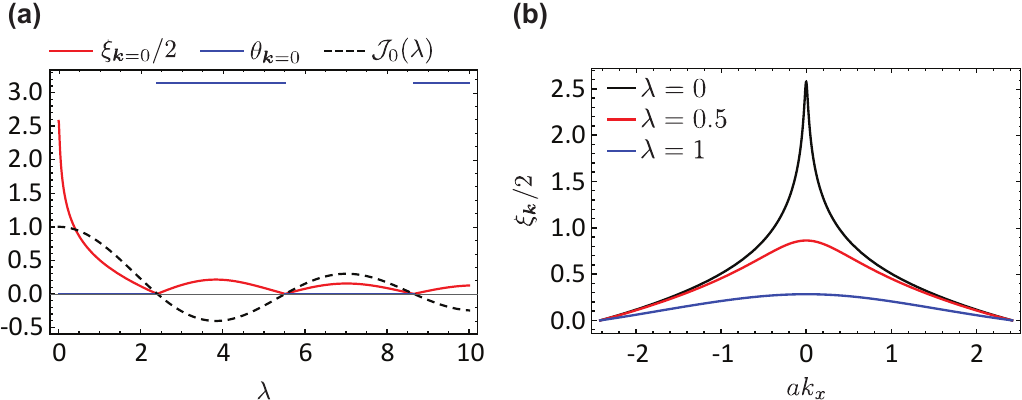}\caption{\label{fig:Bogoliubov_parameters}(a) Depiction of the parameters
defining the Bogoliubov transformation, $\xi_{\boldsymbol{k}}$ and
$\theta_{\boldsymbol{k}}$, as a function of $\lambda$ for the $\boldsymbol{k}=0$
mode. The dashed line represents the Bessel function $\mathcal{J}_{0}(\lambda)$
and serves as a guide to the eye. (b) Representation of the momentum
dependence of $\xi_{\boldsymbol{k}}$ along the $k_{y}=0$ line for
three different values of $\lambda$. The parameters $J=1$meV, $K=10^{-4}J$
and $\hbar\omega_{l}=10J$ were used.}
\end{figure}
In Fig. \ref{fig:Bogoliubov_parameters}a we depict $\xi_{\boldsymbol{k}}$
for the $\boldsymbol{k}=0$ mode, as a function of $\lambda$. Since
we are considering only the isotropic magnon mode, the DMI vanishes
automatically, i.e. $D_{\boldsymbol{k}=0}=0$. For $\lambda=0$, that
is, when the laser field is turned off, we find a large value for
$\xi_{\boldsymbol{k}=0}$; in fact, its value would be even larger
if the easy axis anisotropy had been ignored, since in that case the
squeezing parameter would diverge in the limit $\boldsymbol{k}\rightarrow0$.
As $\lambda$ increases the squeezing parameter decreases, until it
vanishes when $\lambda$ reaches the first zero of the Bessel function
$\mathcal{J}_{0}$. Afterwards, we find that $\xi_{\boldsymbol{k}=0}$
increases slightly and an oscillatory behaviour sets in. It attains
successive maxima when maximum or minimum values of $\mathcal{J}_{0}$
are hit, and vanishes at the zeros of the Bessel function. The maxima
of $\xi_{\boldsymbol{k}=0}$ become progressively smaller as $\lambda$.
Performing an identical analysis for $\boldsymbol{k}\neq0$ one finds
a similar pattern, albeit with an overall smaller magnitude. When
$\boldsymbol{k}\approx\boldsymbol{K}_{\pm}$ we find $\xi_{\boldsymbol{k}}\approx0$,
regardless of $\lambda$, since $\phi_{\boldsymbol{k}}\approx0$ near
the vertices of the Brillouin zone. Thus, we see that the magnon squeezing
tends to decrease as the magnon momentum increases, until it vanishes
at the Dirac points, as depicted in Fig. \ref{fig:Bogoliubov_parameters}b.
Also depicted in Fig. \ref{fig:Bogoliubov_parameters}a, is the dependence
of $\theta_{\boldsymbol{k}}$ with $\lambda$ for the mode $\boldsymbol{k}=0$.
When the laser is turned off, i.e. $\lambda=0$, one finds $\theta_{\boldsymbol{k=}0}=0$.
Interestingly, when $\lambda$ crosses the first zero of the Bessel
function $\mathcal{J}_{0}$, the phase $\theta_{\boldsymbol{k}=0}$
jumps to $\pi$. This phase jumping proves to be the crucial ingredient
to enhance the magnon-photon coupling, which we shall discuss in the
next section. For modes with $\boldsymbol{k}\neq0$ other values are
obtained for $\theta_{\boldsymbol{k}}$, in agreement with Eq. (\ref{fig:Bogoliubov_parameters}),
but the $\pi$-phase jumps remain.

\section{Magnon-photon coupling\label{sec:Magnon-photon-coupling}}

In this section we consider that the antiferromagnet is placed inside
an optical cavity, and study how the squeezed magnons couple with
the cavity photons. The dependence of the magnon-photon coupling on
the external laser field will be studied, and the transmission spectrum
through the cavity computed.

\subsection{System Hamiltonian}

The first step to study the interaction of the squeezed magnons with
the cavity photons is to define the Hamiltonian of the system. This
Hamiltonian is composed of three distinct contributions
\begin{equation}
H=H_{\textrm{cav}}+H_{\textrm{AFM}}+H_{\textrm{int}}
\end{equation}
where $H_{\textrm{cav}}=\sum_{\boldsymbol{q}}\hbar\omega_{\boldsymbol{q}}p_{\boldsymbol{q}}^{\dagger}p_{\boldsymbol{q}}$
is the Hamiltonian of the cavity photons, with $p_{\boldsymbol{q}}/p_{\boldsymbol{q}}^{\dagger}$
the annihilation/creation operator of a photon with momentum $\boldsymbol{q}$
and energy $\hbar\omega_{\boldsymbol{q}}=\hbar c|\boldsymbol{q}|$;
$H_{\textrm{AFM}}$ is the effective Hamiltonian we introduced in
the previous section to describe the antiferromagnet under the incidence
of the external laser field, and $H_{\textrm{int}}=g\mu_{B}\sum_{i\in A}\boldsymbol{b}(\boldsymbol{r}_{i})\cdot\boldsymbol{S}_{A}(\boldsymbol{r}_{i})+g\mu_{B}\sum_{i\in B}\boldsymbol{b}(\boldsymbol{r}_{i})\cdot\boldsymbol{S}_{B}(\boldsymbol{r}_{i})$
is the interaction Hamiltonian between the magnetic field of the cavity
photons, $\boldsymbol{b}$, and the spins of the antiferromagnet.

We now emphasize that up to this point all the momenta we considered
were two-dimensional, due to the in-plane nature of the magnons we
are studying. However, since the cavity photons carry a three dimensional
momentum, we must differentiate between the two. To that end, henceforth,
when considering a two-dimensional momentum we shall label it with
the index $||$, indicating its in-plane configuration (e.g. $\boldsymbol{q}_{||}$
corresponds to the in-plane component of the three-dimensional momentum
$\boldsymbol{q}$).

Considering the magnetic field of the cavity photons to be circularly
polarized, we write its quantized form as \citep{harder2016study}
\begin{align}
\boldsymbol{b}^{\pm}\left(\boldsymbol{r}\right)=\frac{1}{c}\sum_{\boldsymbol{q}}\sqrt{\frac{\hbar\omega_{\boldsymbol{q}}}{4\epsilon_{0}V}}\bigg[ & \left(p_{\boldsymbol{q}}e^{i\boldsymbol{q}\cdot\boldsymbol{r}}+p_{\boldsymbol{q}}^{\dagger}e^{-i\boldsymbol{q}\cdot\boldsymbol{r}}\right)\hat{\boldsymbol{x}}\nonumber \\
\pm i & \left(p_{\boldsymbol{q}}e^{i\boldsymbol{q}\cdot\boldsymbol{r}}-p_{\boldsymbol{q}}^{\dagger}e^{-i\boldsymbol{q}\cdot\boldsymbol{r}}\right)\hat{\boldsymbol{y}}\bigg],
\end{align}
where $\pm$ stands for the two possible circular polarizations, $c$
is the speed of light, $\epsilon_{0}$ is the vacuum permittivity
and $V$ the volume of the cavity. To express $H_{\textrm{int}}$
in terms of the eigenmodes of the antiferromagnet, that is, using
the operators $\alpha_{\boldsymbol{k}}$ and $\beta_{\boldsymbol{k}}$,
we start by expressing $\boldsymbol{S}_{A/B}(\boldsymbol{r}_{i})$
in terms of the sublattice magnon operators, $a_{i}$ and $b_{i}$,
through the linearized Holstein-Primakoff relations given in the previous
section. Then, the Fourier components of these operators are introduced,
and the Bogoliubov transformation inverted, in order to express $a_{\boldsymbol{k}}$
and $b_{\boldsymbol{k}}$ in terms of $\alpha_{\boldsymbol{k}}$ and
$\beta_{\boldsymbol{k}}$. At last, dropping terms with the product
of two annihilation or two creation operators, one finds
\begin{align}
H_{\textrm{int}}^{+} & =\sum_{\boldsymbol{q}}\mathcal{U}_{-\boldsymbol{q}}p_{\boldsymbol{q}}\beta_{-\boldsymbol{q}_{||}}^{\dagger}+\textrm{h.c.}\label{eq:H_int_+}\\
H_{\textrm{int}}^{-} & =\sum_{\boldsymbol{q}}\mathcal{U}_{\boldsymbol{q}}p_{\boldsymbol{q}}\alpha_{\boldsymbol{q}_{||}}^{\dagger}+\textrm{h.c.}\label{eq:H_int_-}
\end{align}
where, once again, the superscript $\pm$ refers to the two circular
polarization of the cavity photons, $\boldsymbol{q}_{||}$ refers
to the in-plane component of the 3D-momentum $\boldsymbol{q}$, and
\begin{equation}
\mathcal{U}_{\boldsymbol{q}}=\frac{g\mu_{B}}{c}\sqrt{\frac{NS\hbar\omega_{\boldsymbol{q}}}{2\epsilon_{0}V}}\left(\cosh\frac{\xi_{\boldsymbol{\boldsymbol{q}}_{||}}}{2}-e^{i\theta_{\boldsymbol{\boldsymbol{q}}_{||}}}\sinh\frac{\xi_{\boldsymbol{\boldsymbol{q}}_{||}}}{2}\right)
\end{equation}
is the magnon-photon coupling. Notice how according to Eqs. (\ref{eq:H_int_+})
and (\ref{eq:H_int_-}) the two orthogonal circular polarizations
couple selectively with just one of the magnon modes each, with equal
coupling strength. Although in the system we are considering the two
magnon modes are degenerate, the polarization of the cavity photons
could be used to select a specific magnon branch in a system where
said degeneracy is broken. Also, we note that for a fixed energy of
the cavity photons (and thus for a fixed $\boldsymbol{q}$), the value
of $\boldsymbol{q}_{||}$ can be tuned by changing the relative orientation
of the antiferromagnet and the magnetic field of the cavity mode.

\subsection{Coupling strength}

Let us now study in more detail the magnon-photon coupling strength.
The coupling $\mathcal{U}_{\boldsymbol{q}}$ is composed of two distinct
contributions: (i) a numerical pre-factor $\mathcal{A}_{\boldsymbol{q}}=\left(g\mu_{B}/c\right)\sqrt{NS\hbar\omega_{\boldsymbol{q}}/2\epsilon_{0}V}$
determined by the properties of the system, namely the cavity photon
energy, cavity volume and number of spins, and (ii) an additional
multiplicative term determined by the coefficients of the Bogoliubov
transformation, $f_{\boldsymbol{q}_{||}}=\cosh\frac{\xi_{\boldsymbol{\boldsymbol{q}}_{||}}}{2}-e^{i\theta_{\boldsymbol{\boldsymbol{q}}_{||}}}\sinh\frac{\xi_{\boldsymbol{\boldsymbol{q}}_{||}}}{2}$,
which depends on the external laser field. The latter contribution
is the one we are interested in studying, in particular how its modulus
$|f_{\boldsymbol{q}_{||}}|=\sqrt{\cosh\xi_{\boldsymbol{\boldsymbol{q}}_{||}}-\cos\theta_{\boldsymbol{\boldsymbol{q}}_{||}}\sinh\xi_{\boldsymbol{\boldsymbol{q}}_{||}}}$
is affected by the momentum dependence and the applied laser.

Consider first the case where the laser is turned off, $\lambda=0$.
According to Fig. \ref{fig:Bogoliubov_parameters}a, we find $\theta_{\boldsymbol{q}_{||}}=0$,
which leads to $|f_{\boldsymbol{q}_{||}}|=e^{-\xi_{\boldsymbol{q}_{||}}/2}$.
Following the analysis of the previous section regarding the squeezing
parameter, we know that it takes its largest value for $\boldsymbol{q}_{||}=0$
and monotonically decreases as the momentum increases, until it vanishes
at the Dirac points (where the magnons are no longer squeezed). Hence,
in the absence of the external laser, we find that the magnon-photon
coupling decreases exponentially as the magnon momentum approaches
the center of the Brillouin zone. In particular, for the parameters
$J=1$meV, $K=10^{-4}J$ and $\hbar\omega_{l}=10J$, we find $|f_{\boldsymbol{q}_{||}=0}|/|f_{\boldsymbol{q}_{||}=\boldsymbol{K}_{\pm}}|\approx0.05$,
that is, the magnon-photon coupling is 20 times smaller near the Brillouin
zone center than at the Dirac points.
\begin{figure}[h]
\centering{}\includegraphics[scale=0.8]{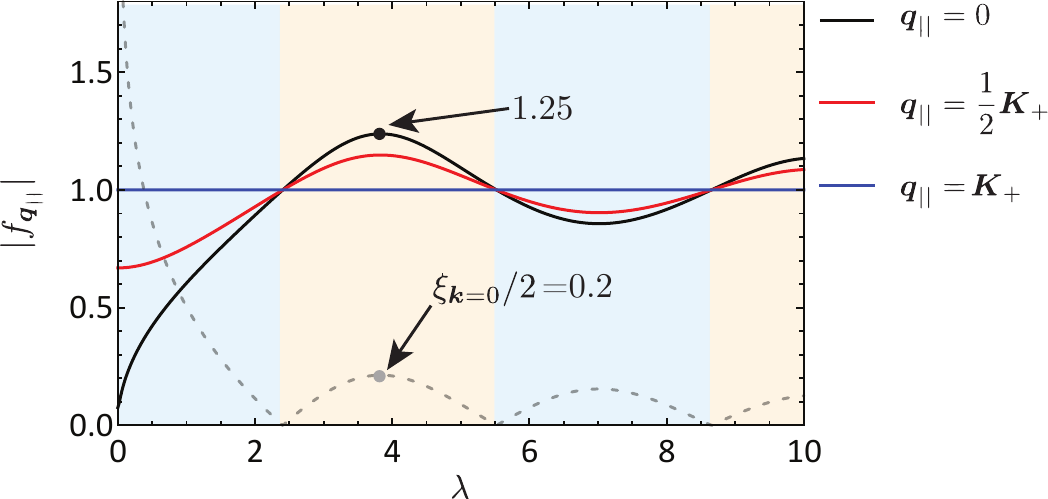}\caption{\label{fig:magnon-photon coupling}Plot of $|f_{\boldsymbol{q}_{||}}|=\sqrt{\cosh\xi_{\boldsymbol{\boldsymbol{q}}_{||}}-\cos\theta_{\boldsymbol{\boldsymbol{q}}_{||}}\sinh\xi_{\boldsymbol{\boldsymbol{q}}_{||}}}$
as a function of $\lambda$ for three different momenta, $\boldsymbol{q}_{||}=0$,
$\boldsymbol{q}_{||}=\boldsymbol{K}_{+}/2$ and $\boldsymbol{q}_{||}=3\boldsymbol{K}_{+}/4$,
with $\boldsymbol{K}_{+}=\left(4\pi/3\sqrt{3}a,0\right)$. The dashed
line shows the the evolution of $\xi_{\boldsymbol{\boldsymbol{q}}_{||}=0}$
with $\lambda$ (also shown in Fig. \ref{fig:Bogoliubov_parameters}a).
The areas shaded in blue/orange refer to the range of $\lambda$ where
the magnon-photon coupling is below/above its value at the Dirac points
when the laser is absent. The parameters $J=1$meV, $K=10^{-4}J$
and $\hbar\omega_{l}=10J$ were used.}
\end{figure}

Let us now consider that $\lambda\neq0$, corresponding to the scenario
where the laser is turned on. Initially we have $\theta_{\boldsymbol{q}_{||}}=0$,
and as $\lambda$ increases the squeezing parameter $\xi_{\boldsymbol{q}_{||}}$
decreases, leading to an increase of the coupling strength $|f_{\boldsymbol{q}_{||}}|$
when compared to the case where the laser is absent. When $\lambda$
reaches the first zero of the Bessel function $\mathcal{J}_{0}$,
the squeezing parameter vanishes on the entire Brillouin zone and
$|f_{\boldsymbol{q}_{||}}|=1\forall\boldsymbol{q}_{||}\in\textrm{1BZ}$.
We note, however, that a similar effect could be achieved without
the laser by increasing the easy axis anisotropy, which for a large
enough value would yield a similar result. The unique feature introduced
by the laser arises when the system is driven in such a way that $\lambda$
is located between the first two zeroes of the $\mathcal{J}_{0}$.
In that situation, and focusing on the $\boldsymbol{q}_{||}=0$ mode,
we find $\theta_{\boldsymbol{q}=0}=\pi$, leading to a coupling strength
which grows exponentially with $\xi_{\boldsymbol{q}_{||}}$, that
is $|f_{\boldsymbol{q}_{||}}|=e^{\xi_{\boldsymbol{q}_{||}}/2}$. Since
in the considered range of values for $\lambda$ we have $\xi_{\boldsymbol{q}}>0$,
then $|f_{\boldsymbol{q}_{||}=0}|>1$. In fact, one finds that $|f_{\boldsymbol{q}_{||}=0}|$
reaches a peak of approximately 1.25 when $\lambda\approx3.8$ (corresponding
to the first minimum of $\mathcal{J}_{0}$). The peak value $|f_{\boldsymbol{q}_{||}=0}|\approx1.25$
is obtained in the limit where the exchange coupling $J$ dominates
the anisotropy term $K$. If $K$ becomes comparable with $J$, then
the maximum value for $|f_{\boldsymbol{q}_{||}=0}|$ decreases.

Hence, by driving the system with a laser field, we are able to enhance
the coupling of the $\boldsymbol{q}_{||}=0$ magnons with the cavity
photons by a factor of approximately 25 when compared to the case
where no laser is applied. Not only that, but we are also able to
push the coupling strength beyond the limit of what is found for unsqueezed
magnons, whilst retaining some magnon squeezing ($\xi_{\boldsymbol{q}_{||}=0}\approx0.2$
for $\lambda\approx3.8$); a result only attainable by driving the
system with the laser field. For magnon modes with $\boldsymbol{q}_{||}\neq0$
the laser induced enhanced coupling is also present, although the
effect becomes progressively weaker as we approach the Dirac points.
These results are summarized in Fig. \ref{fig:magnon-photon coupling}
where we depict $|f_{\boldsymbol{q}_{||}}|$ as a function of $\lambda$
for different momenta $\boldsymbol{q}_{||}$. The shaded blue/orange
areas indicate the range of $\lambda$ where the magnon-photon coupling
is smaller/larger than for unsqueezed magnons.

\subsection{Transmission spectra}

Now that the Hamiltonian of the system has been determined, and the
details of the magnon-photon coupling have been discussed, we move
on to the computation of the transmission spectra through the cavity.
Following \citep{harder2018cavity}, we state that the transmission
through the cavity should be proportional to the spectral function
of the cavity photons, which follows directly from the retarded Green's
function $G^{R}(t)=-\frac{i}{\hbar}\theta(t)\langle p_{\boldsymbol{q}}(t)p_{\boldsymbol{q}}^{\dagger}(0)\rangle$.
Considering now that the cavity has only one mode, and that its magnetic
field is polarized with the ``$-$'' circular polarization, we obtain
the following expression for the retarded Green's function \citep{harder2018cavity,MahanManyParticle}
\begin{equation}
G^{R}(\hbar\omega)=\frac{1}{\hbar\omega-\hbar\omega_{\boldsymbol{q}}+i\Gamma-\frac{|\mathcal{U}_{\boldsymbol{q}}|^{2}}{\hbar\omega-\epsilon_{\boldsymbol{q}_{||}}+i\delta}},
\end{equation}
where $\Gamma$ and $\delta$ are the cavity and antiferromagnet intrinsic
damping factors, and $\hbar\omega$ is the energy of the incident
photons. Once more we have $\hbar\omega_{\boldsymbol{q}}=\hbar c|\boldsymbol{q}|$
the energy of the cavity mode and $\epsilon_{\boldsymbol{q}_{||}}$
the energy of the magnon mode that couples with the cavity photons.
Although the circular polarization we chose only allows coupling to
the $\alpha-$magnons, Eq. (\ref{eq:H_int_-}), a similar result would
be obtained for the other circular polarization where coupling to
$\beta-$magnons would appear instead. Since for the system we are
considering the two magnon modes are degenerate, the transmission
spectrum is identical in both cases.

Having determined $G^{R}(\hbar\omega_{\boldsymbol{q}})$ we define
the transmission amplitude as $t(\hbar\omega)\propto-2\textrm{Im}G^{R}(\hbar\omega)$.
A simple way to avoid the proportionality relation, is to normalize
the transmission amplitude by its value when the antiferromagnet is
not present in the cavity ($\mathcal{U}_{\boldsymbol{q}}=0$) and
the incident photons are in resonance with the cavity mode $\hbar\omega=\hbar\omega_{\boldsymbol{q}}$.
Doing so, we find
\begin{equation}
\bar{t}(\hbar\omega)=-\textrm{Im}\frac{\Gamma}{\hbar\omega-\hbar\omega_{\boldsymbol{q}}+i\Gamma-\frac{|\mathcal{U}_{\boldsymbol{q}}|^{2}}{\hbar\omega-\epsilon_{\boldsymbol{q}_{||}}+i\delta}},
\end{equation}
where $\bar{t}(\hbar\omega)$ is the normalized transmission amplitude.
The poles of $\bar{t}(\hbar\omega)$ correspond to the excitations
of the system, which are termed magnon-polaritons as they correspond
to the hybridization of the antiferromagnet magnons with the cavity
photons.
\begin{figure}[h]
\centering{}\includegraphics[scale=0.9]{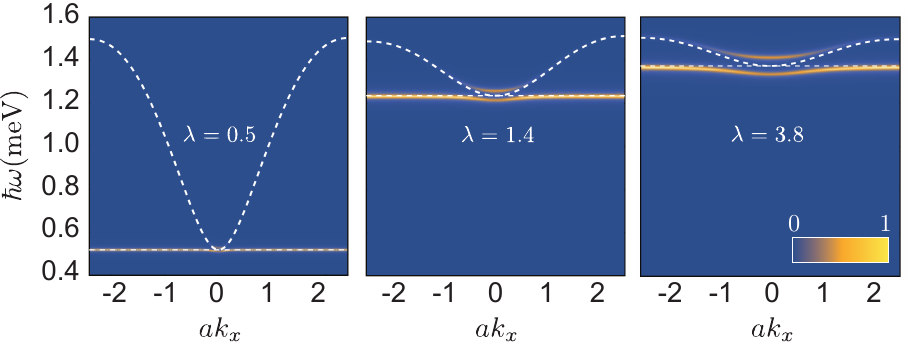}\caption{\label{fig:t(w)}Density plots of the normalized transmission $\bar{t}(\hbar\omega)$
as a function of the incident photon's energy and the in-plane magnon
momentum (along the line with $k_{y}=0$). In each plot a different
value for $\lambda$ was considered and the energy of the cavity mode
(horizontal dashed line) was chosen to match the minimum of the magnon
dispersion (curved dashed line). The parameters $J=1$meV, $K=10^{-4}J$
and $\hbar\omega_{l}=50J$ were used. We also considered $\Gamma=\delta=10^{-3}J$,
and $\mathcal{A}_{\boldsymbol{q}}=8\times10^{-4}\sqrt{\hbar\omega_{\boldsymbol{q}}}$
(in eV).}
\end{figure}

In Fig. \ref{fig:t(w)} we depict a density plot of $\bar{t}(\hbar\omega)$
as a function of the incident photon energy $\hbar\omega$ and the
magnon momentum $\boldsymbol{k}_{||}$, along the line $k_{y}=0$
on the first Brillouin zone, for three different values of $\lambda$;
for each case the energy of the cavity mode was chosen to match the
minimum of the magnon dispersion. The first thing we note regarding
this result is that the bandwidth of the magnon dispersion is modified
by changing $\lambda$ in agreement with what was previously found
in Fig. \ref{fig:magnon_dispersion} (the effect of the DMI is small
since we are considering $\hbar\omega_{l}=50J$). Focusing on the
leftmost panel, we see that the hybridization between the cavity mode
and the zero-momentum magnon is almost nonexistent; this is simultaneously
due to the small energy of the cavity photon (leading to a small $\mathcal{A}_{\boldsymbol{k}}$),
and due to the fact that $|f_{\boldsymbol{k}_{||}}|$ is small for
$\lambda=0.5$. Regarding the middle panel, we see that, due to the
higher value used for $\lambda$, the bottom of the magnon dispersion
moves to higher energies, and, at the same time, the magnitude of
$|f_{\boldsymbol{k}_{||}}|$ increases (see Fig. \ref{fig:magnon-photon coupling}).
Because of this, one finds a much clearer coupling between the magnons
and the cavity photons, which is identified by the anticrossing of
the magnon-polariton branches. At last, on the rightmost panel, where
$\lambda=3.8$, we observe the largest hybridization between magnons
and photons, with a clear separation between the two magnon-polariton
modes. This strong coupling was expected, since $\lambda$ was carefully
chosen to maximize $|f_{\boldsymbol{k}_{||}}|$, in agreement with
the discussion following Fig. \ref{fig:magnon-photon coupling}.

\section{Plausible parameters\label{sec:Plausible-parameters}}

Up to this point we have discussed the general physical properties
of a laser driven antiferromagnet without focusing on a particular
implementation of the system. In this section we wish to estimate
the values the different parameters the system should have in order
to observe the phenomena discussed so far. To the best of our knowledge, an experiment focused on the phenomena we describe in this paper has not been performed so far.

To do so, we start by setting $\lambda=2.4$, corresponding to the
minimum value of $\lambda$ which allows an enhanced magnon-photon
coupling above the unsqueezed limit (see Fig. \ref{fig:magnon-photon coupling}).
Recalling $\lambda=g\mu_{B}E_{0}a/\hbar c^{2}$, and setting $g=2$,
we find that in order to have $\lambda=2.4$ we must have $E_{0}a\approx1.2\times10^{6}$
V. For $a=3.3\text{Å}$ \citep{gong2018electrically}, corresponding
to a value in the order of magnitude that is usually found for hexagonal
boron nitride, transition metal dichalcogenides, etc., we obtain $E_{0}\approx3\times10^{13}\textrm{V/cm}$.
If a larger $a$ is used, a smaller electric field is required to
produce the same $\lambda$. In Ref. \citep{yoon2021realization}
a laser intensity of $10^{23}\textrm{W/cm}^{2}$ has been reported,
which in terms of the electric field's magnitude corresponds to roughly
$E_{0}\approx9\times10^{12}\textrm{V/cm}$, within the order of magnitude
of the required value. Hence, current state of the art laser technology
is already compatible with the requirements to significantly increase
the magnon-photon coupling. To estimate the numerical value of $\mathcal{U}_{\boldsymbol{q}}$
we focus on the magnitude of $\mathcal{A}_{\boldsymbol{q}}=\left(g\mu_{B}/c\right)\sqrt{NS\hbar\omega_{\boldsymbol{q}}/2\epsilon_{0}V}$.
We now set $S=1/2$ and note that $N/V=\rho/2h$ where $\rho/2$ is
the density of spins and $h$ is the height of the cavity. Considering
$h=40\textrm{nm}$ \citep{noginov2009demonstration} and $\rho/2\sim10^{14}\textrm{cm}^{-2}$
(in agreement with the measured values of \citep{gong2018electrically},
and compatible with the value of $a$ considered above), we find $\mathcal{A}_{\boldsymbol{q}}\approx10^{-4}\sqrt{\hbar\omega_{\boldsymbol{q}}}$
(in eV), which is comparable with what we used in Fig. \ref{fig:t(w)}.
For $\hbar\omega\sim1\textrm{meV}$ we find $\mathcal{A}_{\boldsymbol{q}}\sim3\times10^{-6}$
eV, corresponding to approximately 0.7 $\textrm{GHz}$, which is well
within the resolution of current experiments \citep{harder2016study,harder2018level,yao2019microscopic}.

\section{Final remarks\label{sec:Final-remarks}}

In this paper we studied a honeycomb antiferromagnet subject to an
external laser field. After a Floquet effective Hamiltonian was introduced,
we diagonalized it using a Bogoliubov transformation, and found that
the eigenexcitations of the system correspond to squeezed magnons.
The presence of the laser allows the control of several properties
of these spin excitations, namely their dispersion relation, and squeezing
parameters.

When studying how the magnons couple with photons in a cavity, we
found that the two magnon modes of the system couple selectively with
the two orthogonal circular polarizations. Furthermore, we found that
by tuning the intensity of the applied laser field the strength of
the magnon-photon coupling can be significantly enhanced. In particular,
we found that, for the isotropic magnon mode, the coupling strength
with the cavity photons can be enhanced by an order of magnitude when
compared to what is found in the absence of the laser. When the system
is not driven, the largest magnon-photon coupling is achieved at the
Dirac points, where the magnons do not present squeezing. When the
laser is present, however, the maximum coupling is reached for the
magnons near the Brillouin zone center, which preserve some of the
original squeezing; in this case the magnon-photon coupling surpasses
the maximum value found for the undriven system.

To the best of our knowledge, an experiment probing these physical effects has not yet been realized. However, from our estimates, state of the art equipment may be able to probe the discussed phenomena. Since our theoretical predictions were derived for a strictly 2D monolayer, the results may differ if experiments are realized on thin films, due to the sensitivity of these materials regarding the number of layers.

Finally, we note that although we focused on the magnon-photon interaction,
other options could be explored. An example of this is the interaction
of magnons with conduction electrons. According to Ref. \citep{kamra2019antiferromagnetic}
the transition rate for the electron-magnon scattering takes a similar
form to the function $f_{\boldsymbol{q}_{||}}$ we identified in the
magnon-photon coupling. Due to this similarity, the ideas discussed
here regarding the coupling strength could, in principle, also be
applied in that type of system.
\begin{acknowledgments}
J.C.G.H. acknowledges the Center of Physics for a grant funded by
the UIDB/04650/2020 strategic project. T. V. C. A. acknowledges funding
through the project QML-HEP - CERN/FIS-COM/0004/2021 . N. M. R. P.
acknowledges financial support from FCT (grant PTDC/FIS-MAC/2045/2021)
and from the European Commission through the project GrapheneDriven
Revolutions in ICT and Beyond (Ref. No. 881603, CORE 3).
\end{acknowledgments}

\bibliographystyle{apsrev4-2}

\begin{thebibliography}{51}%
\makeatletter
\providecommand \@ifxundefined [1]{%
 \@ifx{#1\undefined}
}%
\providecommand \@ifnum [1]{%
 \ifnum #1\expandafter \@firstoftwo
 \else \expandafter \@secondoftwo
 \fi
}%
\providecommand \@ifx [1]{%
 \ifx #1\expandafter \@firstoftwo
 \else \expandafter \@secondoftwo
 \fi
}%
\providecommand \natexlab [1]{#1}%
\providecommand \enquote  [1]{``#1''}%
\providecommand \bibnamefont  [1]{#1}%
\providecommand \bibfnamefont [1]{#1}%
\providecommand \citenamefont [1]{#1}%
\providecommand \href@noop [0]{\@secondoftwo}%
\providecommand \href [0]{\begingroup \@sanitize@url \@href}%
\providecommand \@href[1]{\@@startlink{#1}\@@href}%
\providecommand \@@href[1]{\endgroup#1\@@endlink}%
\providecommand \@sanitize@url [0]{\catcode `\\12\catcode `\$12\catcode
  `\&12\catcode `\#12\catcode `\^12\catcode `\_12\catcode `\%12\relax}%
\providecommand \@@startlink[1]{}%
\providecommand \@@endlink[0]{}%
\providecommand \url  [0]{\begingroup\@sanitize@url \@url }%
\providecommand \@url [1]{\endgroup\@href {#1}{\urlprefix }}%
\providecommand \urlprefix  [0]{URL }%
\providecommand \Eprint [0]{\href }%
\providecommand \doibase [0]{https://doi.org/}%
\providecommand \selectlanguage [0]{\@gobble}%
\providecommand \bibinfo  [0]{\@secondoftwo}%
\providecommand \bibfield  [0]{\@secondoftwo}%
\providecommand \translation [1]{[#1]}%
\providecommand \BibitemOpen [0]{}%
\providecommand \bibitemStop [0]{}%
\providecommand \bibitemNoStop [0]{.\EOS\space}%
\providecommand \EOS [0]{\spacefactor3000\relax}%
\providecommand \BibitemShut  [1]{\csname bibitem#1\endcsname}%
\let\auto@bib@innerbib\@empty
\bibitem [{\citenamefont {Chumak}\ \emph {et~al.}(2015)\citenamefont {Chumak},
  \citenamefont {Vasyuchka}, \citenamefont {Serga},\ and\ \citenamefont
  {Hillebrands}}]{chumak2015magnon}%
  \BibitemOpen
  \bibfield  {author} {\bibinfo {author} {\bibfnamefont {A.~V.}\ \bibnamefont
  {Chumak}}, \bibinfo {author} {\bibfnamefont {V.~I.}\ \bibnamefont
  {Vasyuchka}}, \bibinfo {author} {\bibfnamefont {A.~A.}\ \bibnamefont
  {Serga}},\ and\ \bibinfo {author} {\bibfnamefont {B.}~\bibnamefont
  {Hillebrands}},\ }\href@noop {} {\bibfield  {journal} {\bibinfo  {journal}
  {Nature Physics}\ }\textbf {\bibinfo {volume} {11}},\ \bibinfo {pages} {453}
  (\bibinfo {year} {2015})}\BibitemShut {NoStop}%
\bibitem [{\citenamefont {Yuan}\ \emph {et~al.}(2021)\citenamefont {Yuan},
  \citenamefont {Cao}, \citenamefont {Kamra}, \citenamefont {Duine},\ and\
  \citenamefont {Yan}}]{yuan2021quantum}%
  \BibitemOpen
  \bibfield  {author} {\bibinfo {author} {\bibfnamefont {H.}~\bibnamefont
  {Yuan}}, \bibinfo {author} {\bibfnamefont {Y.}~\bibnamefont {Cao}}, \bibinfo
  {author} {\bibfnamefont {A.}~\bibnamefont {Kamra}}, \bibinfo {author}
  {\bibfnamefont {R.~A.}\ \bibnamefont {Duine}},\ and\ \bibinfo {author}
  {\bibfnamefont {P.}~\bibnamefont {Yan}},\ }\href@noop {} {\bibfield
  {journal} {\bibinfo  {journal} {arXiv preprint arXiv:2111.14241}\ } (\bibinfo
  {year} {2021})}\BibitemShut {NoStop}%
\bibitem [{\citenamefont {Chumak}\ \emph {et~al.}(2014)\citenamefont {Chumak},
  \citenamefont {Serga},\ and\ \citenamefont {Hillebrands}}]{chumak2014magnon}%
  \BibitemOpen
  \bibfield  {author} {\bibinfo {author} {\bibfnamefont {A.~V.}\ \bibnamefont
  {Chumak}}, \bibinfo {author} {\bibfnamefont {A.~A.}\ \bibnamefont {Serga}},\
  and\ \bibinfo {author} {\bibfnamefont {B.}~\bibnamefont {Hillebrands}},\
  }\href@noop {} {\bibfield  {journal} {\bibinfo  {journal} {Nature
  communications}\ }\textbf {\bibinfo {volume} {5}},\ \bibinfo {pages} {1}
  (\bibinfo {year} {2014})}\BibitemShut {NoStop}%
\bibitem [{\citenamefont {Demidov}\ \emph {et~al.}(2011)\citenamefont
  {Demidov}, \citenamefont {Kostylev}, \citenamefont {Rott}, \citenamefont
  {M{\"u}nchenberger}, \citenamefont {Reiss},\ and\ \citenamefont
  {Demokritov}}]{demidov2011excitation}%
  \BibitemOpen
  \bibfield  {author} {\bibinfo {author} {\bibfnamefont {V.}~\bibnamefont
  {Demidov}}, \bibinfo {author} {\bibfnamefont {M.}~\bibnamefont {Kostylev}},
  \bibinfo {author} {\bibfnamefont {K.}~\bibnamefont {Rott}}, \bibinfo {author}
  {\bibfnamefont {J.}~\bibnamefont {M{\"u}nchenberger}}, \bibinfo {author}
  {\bibfnamefont {G.}~\bibnamefont {Reiss}},\ and\ \bibinfo {author}
  {\bibfnamefont {S.}~\bibnamefont {Demokritov}},\ }\href@noop {} {\bibfield
  {journal} {\bibinfo  {journal} {Applied Physics Letters}\ }\textbf {\bibinfo
  {volume} {99}},\ \bibinfo {pages} {082507} (\bibinfo {year}
  {2011})}\BibitemShut {NoStop}%
\bibitem [{\citenamefont {Khitun}\ \emph {et~al.}(2010)\citenamefont {Khitun},
  \citenamefont {Bao},\ and\ \citenamefont {Wang}}]{khitun2010magnonic}%
  \BibitemOpen
  \bibfield  {author} {\bibinfo {author} {\bibfnamefont {A.}~\bibnamefont
  {Khitun}}, \bibinfo {author} {\bibfnamefont {M.}~\bibnamefont {Bao}},\ and\
  \bibinfo {author} {\bibfnamefont {K.~L.}\ \bibnamefont {Wang}},\ }\href@noop
  {} {\bibfield  {journal} {\bibinfo  {journal} {Journal of Physics D: Applied
  Physics}\ }\textbf {\bibinfo {volume} {43}},\ \bibinfo {pages} {264005}
  (\bibinfo {year} {2010})}\BibitemShut {NoStop}%
\bibitem [{\citenamefont {Schneider}\ \emph {et~al.}(2008)\citenamefont
  {Schneider}, \citenamefont {Serga}, \citenamefont {Leven}, \citenamefont
  {Hillebrands}, \citenamefont {Stamps},\ and\ \citenamefont
  {Kostylev}}]{schneider2008realization}%
  \BibitemOpen
  \bibfield  {author} {\bibinfo {author} {\bibfnamefont {T.}~\bibnamefont
  {Schneider}}, \bibinfo {author} {\bibfnamefont {A.~A.}\ \bibnamefont
  {Serga}}, \bibinfo {author} {\bibfnamefont {B.}~\bibnamefont {Leven}},
  \bibinfo {author} {\bibfnamefont {B.}~\bibnamefont {Hillebrands}}, \bibinfo
  {author} {\bibfnamefont {R.~L.}\ \bibnamefont {Stamps}},\ and\ \bibinfo
  {author} {\bibfnamefont {M.~P.}\ \bibnamefont {Kostylev}},\ }\href@noop {}
  {\bibfield  {journal} {\bibinfo  {journal} {Applied Physics Letters}\
  }\textbf {\bibinfo {volume} {92}},\ \bibinfo {pages} {022505} (\bibinfo
  {year} {2008})}\BibitemShut {NoStop}%
\bibitem [{\citenamefont {Soykal}\ and\ \citenamefont
  {Flatt{\'e}}(2010)}]{soykal2010strong}%
  \BibitemOpen
  \bibfield  {author} {\bibinfo {author} {\bibfnamefont {{\"O}.~O.}\
  \bibnamefont {Soykal}}\ and\ \bibinfo {author} {\bibfnamefont
  {M.}~\bibnamefont {Flatt{\'e}}},\ }\href@noop {} {\bibfield  {journal}
  {\bibinfo  {journal} {Physical review letters}\ }\textbf {\bibinfo {volume}
  {104}},\ \bibinfo {pages} {077202} (\bibinfo {year} {2010})}\BibitemShut
  {NoStop}%
\bibitem [{\citenamefont {Harder}\ \emph {et~al.}(2016)\citenamefont {Harder},
  \citenamefont {Bai}, \citenamefont {Match}, \citenamefont {Sirker},\ and\
  \citenamefont {Hu}}]{harder2016study}%
  \BibitemOpen
  \bibfield  {author} {\bibinfo {author} {\bibfnamefont {M.}~\bibnamefont
  {Harder}}, \bibinfo {author} {\bibfnamefont {L.}~\bibnamefont {Bai}},
  \bibinfo {author} {\bibfnamefont {C.}~\bibnamefont {Match}}, \bibinfo
  {author} {\bibfnamefont {J.}~\bibnamefont {Sirker}},\ and\ \bibinfo {author}
  {\bibfnamefont {C.}~\bibnamefont {Hu}},\ }\href@noop {} {\bibfield  {journal}
  {\bibinfo  {journal} {Science China Physics, Mechanics \& Astronomy}\
  }\textbf {\bibinfo {volume} {59}},\ \bibinfo {pages} {1} (\bibinfo {year}
  {2016})}\BibitemShut {NoStop}%
\bibitem [{\citenamefont {Harder}\ \emph {et~al.}(2021)\citenamefont {Harder},
  \citenamefont {Yao}, \citenamefont {Gui},\ and\ \citenamefont
  {Hu}}]{harder2021coherent}%
  \BibitemOpen
  \bibfield  {author} {\bibinfo {author} {\bibfnamefont {M.}~\bibnamefont
  {Harder}}, \bibinfo {author} {\bibfnamefont {B.}~\bibnamefont {Yao}},
  \bibinfo {author} {\bibfnamefont {Y.}~\bibnamefont {Gui}},\ and\ \bibinfo
  {author} {\bibfnamefont {C.-M.}\ \bibnamefont {Hu}},\ }\href@noop {}
  {\bibfield  {journal} {\bibinfo  {journal} {Journal of Applied Physics}\
  }\textbf {\bibinfo {volume} {129}},\ \bibinfo {pages} {201101} (\bibinfo
  {year} {2021})}\BibitemShut {NoStop}%
\bibitem [{\citenamefont {Tabuchi}\ \emph {et~al.}(2015)\citenamefont
  {Tabuchi}, \citenamefont {Ishino}, \citenamefont {Noguchi}, \citenamefont
  {Ishikawa}, \citenamefont {Yamazaki}, \citenamefont {Usami},\ and\
  \citenamefont {Nakamura}}]{tabuchi2015coherent}%
  \BibitemOpen
  \bibfield  {author} {\bibinfo {author} {\bibfnamefont {Y.}~\bibnamefont
  {Tabuchi}}, \bibinfo {author} {\bibfnamefont {S.}~\bibnamefont {Ishino}},
  \bibinfo {author} {\bibfnamefont {A.}~\bibnamefont {Noguchi}}, \bibinfo
  {author} {\bibfnamefont {T.}~\bibnamefont {Ishikawa}}, \bibinfo {author}
  {\bibfnamefont {R.}~\bibnamefont {Yamazaki}}, \bibinfo {author}
  {\bibfnamefont {K.}~\bibnamefont {Usami}},\ and\ \bibinfo {author}
  {\bibfnamefont {Y.}~\bibnamefont {Nakamura}},\ }\href@noop {} {\bibfield
  {journal} {\bibinfo  {journal} {Science}\ }\textbf {\bibinfo {volume}
  {349}},\ \bibinfo {pages} {405} (\bibinfo {year} {2015})}\BibitemShut
  {NoStop}%
\bibitem [{\citenamefont {Liu}\ \emph {et~al.}(2019)\citenamefont {Liu},
  \citenamefont {Xiong},\ and\ \citenamefont {Wu}}]{liu2019magnon}%
  \BibitemOpen
  \bibfield  {author} {\bibinfo {author} {\bibfnamefont {Z.-X.}\ \bibnamefont
  {Liu}}, \bibinfo {author} {\bibfnamefont {H.}~\bibnamefont {Xiong}},\ and\
  \bibinfo {author} {\bibfnamefont {Y.}~\bibnamefont {Wu}},\ }\href@noop {}
  {\bibfield  {journal} {\bibinfo  {journal} {Physical Review B}\ }\textbf
  {\bibinfo {volume} {100}},\ \bibinfo {pages} {134421} (\bibinfo {year}
  {2019})}\BibitemShut {NoStop}%
\bibitem [{\citenamefont {Bittencourt}\ \emph {et~al.}(2019)\citenamefont
  {Bittencourt}, \citenamefont {Feulner},\ and\ \citenamefont
  {Kusminskiy}}]{bittencourt2019magnon}%
  \BibitemOpen
  \bibfield  {author} {\bibinfo {author} {\bibfnamefont {V.~A.}\ \bibnamefont
  {Bittencourt}}, \bibinfo {author} {\bibfnamefont {V.}~\bibnamefont
  {Feulner}},\ and\ \bibinfo {author} {\bibfnamefont {S.~V.}\ \bibnamefont
  {Kusminskiy}},\ }\href@noop {} {\bibfield  {journal} {\bibinfo  {journal}
  {Physical Review A}\ }\textbf {\bibinfo {volume} {100}},\ \bibinfo {pages}
  {013810} (\bibinfo {year} {2019})}\BibitemShut {NoStop}%
\bibitem [{\citenamefont {Gerry}\ \emph {et~al.}(2005)\citenamefont {Gerry},
  \citenamefont {Knight},\ and\ \citenamefont
  {Knight}}]{gerry2005introductory}%
  \BibitemOpen
  \bibfield  {author} {\bibinfo {author} {\bibfnamefont {C.}~\bibnamefont
  {Gerry}}, \bibinfo {author} {\bibfnamefont {P.}~\bibnamefont {Knight}},\ and\
  \bibinfo {author} {\bibfnamefont {P.~L.}\ \bibnamefont {Knight}},\
  }\href@noop {} {\emph {\bibinfo {title} {Introductory quantum optics}}}\
  (\bibinfo  {publisher} {Cambridge University Press, New York},\ \bibinfo
  {year} {2005})\BibitemShut {NoStop}%
\bibitem [{\citenamefont {Fox}(2006)}]{fox2006quantum}%
  \BibitemOpen
  \bibfield  {author} {\bibinfo {author} {\bibfnamefont {M.}~\bibnamefont
  {Fox}},\ }\href@noop {} {\emph {\bibinfo {title} {Quantum optics: an
  introduction}}},\ Vol.~\bibinfo {volume} {15}\ (\bibinfo  {publisher} {Oxford
  University Press Inc., New York},\ \bibinfo {year} {2006})\BibitemShut
  {NoStop}%
\bibitem [{\citenamefont {Andersen}\ \emph {et~al.}(2016)\citenamefont
  {Andersen}, \citenamefont {Gehring}, \citenamefont {Marquardt},\ and\
  \citenamefont {Leuchs}}]{andersen201630}%
  \BibitemOpen
  \bibfield  {author} {\bibinfo {author} {\bibfnamefont {U.~L.}\ \bibnamefont
  {Andersen}}, \bibinfo {author} {\bibfnamefont {T.}~\bibnamefont {Gehring}},
  \bibinfo {author} {\bibfnamefont {C.}~\bibnamefont {Marquardt}},\ and\
  \bibinfo {author} {\bibfnamefont {G.}~\bibnamefont {Leuchs}},\ }\href@noop {}
  {\bibfield  {journal} {\bibinfo  {journal} {Physica Scripta}\ }\textbf
  {\bibinfo {volume} {91}},\ \bibinfo {pages} {053001} (\bibinfo {year}
  {2016})}\BibitemShut {NoStop}%
\bibitem [{\citenamefont {Aasi}\ \emph {et~al.}(2013)\citenamefont {Aasi},
  \citenamefont {Abadie}, \citenamefont {Abbott}, \citenamefont {Abbott},
  \citenamefont {Abbott}, \citenamefont {Abernathy}, \citenamefont {Adams},
  \citenamefont {Adams}, \citenamefont {Addesso}, \citenamefont {Adhikari}
  \emph {et~al.}}]{aasi2013enhanced}%
  \BibitemOpen
  \bibfield  {author} {\bibinfo {author} {\bibfnamefont {J.}~\bibnamefont
  {Aasi}}, \bibinfo {author} {\bibfnamefont {J.}~\bibnamefont {Abadie}},
  \bibinfo {author} {\bibfnamefont {B.}~\bibnamefont {Abbott}}, \bibinfo
  {author} {\bibfnamefont {R.}~\bibnamefont {Abbott}}, \bibinfo {author}
  {\bibfnamefont {T.}~\bibnamefont {Abbott}}, \bibinfo {author} {\bibfnamefont
  {M.}~\bibnamefont {Abernathy}}, \bibinfo {author} {\bibfnamefont
  {C.}~\bibnamefont {Adams}}, \bibinfo {author} {\bibfnamefont
  {T.}~\bibnamefont {Adams}}, \bibinfo {author} {\bibfnamefont
  {P.}~\bibnamefont {Addesso}}, \bibinfo {author} {\bibfnamefont
  {R.}~\bibnamefont {Adhikari}}, \emph {et~al.},\ }\href@noop {} {\bibfield
  {journal} {\bibinfo  {journal} {Nature Photonics}\ }\textbf {\bibinfo
  {volume} {7}},\ \bibinfo {pages} {613} (\bibinfo {year} {2013})}\BibitemShut
  {NoStop}%
\bibitem [{\citenamefont {Walls}(1983)}]{walls1983squeezed}%
  \BibitemOpen
  \bibfield  {author} {\bibinfo {author} {\bibfnamefont {D.~F.}\ \bibnamefont
  {Walls}},\ }\href@noop {} {\bibfield  {journal} {\bibinfo  {journal}
  {nature}\ }\textbf {\bibinfo {volume} {306}},\ \bibinfo {pages} {141}
  (\bibinfo {year} {1983})}\BibitemShut {NoStop}%
\bibitem [{\citenamefont {Kamra}\ \emph {et~al.}(2019)\citenamefont {Kamra},
  \citenamefont {Thingstad}, \citenamefont {Rastelli}, \citenamefont {Duine},
  \citenamefont {Brataas}, \citenamefont {Belzig},\ and\ \citenamefont
  {Sudb{\o}}}]{kamra2019antiferromagnetic}%
  \BibitemOpen
  \bibfield  {author} {\bibinfo {author} {\bibfnamefont {A.}~\bibnamefont
  {Kamra}}, \bibinfo {author} {\bibfnamefont {E.}~\bibnamefont {Thingstad}},
  \bibinfo {author} {\bibfnamefont {G.}~\bibnamefont {Rastelli}}, \bibinfo
  {author} {\bibfnamefont {R.~A.}\ \bibnamefont {Duine}}, \bibinfo {author}
  {\bibfnamefont {A.}~\bibnamefont {Brataas}}, \bibinfo {author} {\bibfnamefont
  {W.}~\bibnamefont {Belzig}},\ and\ \bibinfo {author} {\bibfnamefont
  {A.}~\bibnamefont {Sudb{\o}}},\ }\href@noop {} {\bibfield  {journal}
  {\bibinfo  {journal} {Physical Review B}\ }\textbf {\bibinfo {volume}
  {100}},\ \bibinfo {pages} {174407} (\bibinfo {year} {2019})}\BibitemShut
  {NoStop}%
\bibitem [{\citenamefont {Kamra}\ and\ \citenamefont
  {Belzig}(2016)}]{kamra2016super}%
  \BibitemOpen
  \bibfield  {author} {\bibinfo {author} {\bibfnamefont {A.}~\bibnamefont
  {Kamra}}\ and\ \bibinfo {author} {\bibfnamefont {W.}~\bibnamefont {Belzig}},\
  }\href@noop {} {\bibfield  {journal} {\bibinfo  {journal} {Physical review
  letters}\ }\textbf {\bibinfo {volume} {116}},\ \bibinfo {pages} {146601}
  (\bibinfo {year} {2016})}\BibitemShut {NoStop}%
\bibitem [{\citenamefont {Liensberger}\ \emph {et~al.}(2019)\citenamefont
  {Liensberger}, \citenamefont {Kamra}, \citenamefont {Maier-Flaig},
  \citenamefont {Gepr{\"a}gs}, \citenamefont {Erb}, \citenamefont
  {Goennenwein}, \citenamefont {Gross}, \citenamefont {Belzig}, \citenamefont
  {Huebl},\ and\ \citenamefont {Weiler}}]{liensberger2019exchange}%
  \BibitemOpen
  \bibfield  {author} {\bibinfo {author} {\bibfnamefont {L.}~\bibnamefont
  {Liensberger}}, \bibinfo {author} {\bibfnamefont {A.}~\bibnamefont {Kamra}},
  \bibinfo {author} {\bibfnamefont {H.}~\bibnamefont {Maier-Flaig}}, \bibinfo
  {author} {\bibfnamefont {S.}~\bibnamefont {Gepr{\"a}gs}}, \bibinfo {author}
  {\bibfnamefont {A.}~\bibnamefont {Erb}}, \bibinfo {author} {\bibfnamefont
  {S.~T.}\ \bibnamefont {Goennenwein}}, \bibinfo {author} {\bibfnamefont
  {R.}~\bibnamefont {Gross}}, \bibinfo {author} {\bibfnamefont
  {W.}~\bibnamefont {Belzig}}, \bibinfo {author} {\bibfnamefont
  {H.}~\bibnamefont {Huebl}},\ and\ \bibinfo {author} {\bibfnamefont
  {M.}~\bibnamefont {Weiler}},\ }\href@noop {} {\bibfield  {journal} {\bibinfo
  {journal} {Physical review letters}\ }\textbf {\bibinfo {volume} {123}},\
  \bibinfo {pages} {117204} (\bibinfo {year} {2019})}\BibitemShut {NoStop}%
\bibitem [{\citenamefont {Zou}\ \emph {et~al.}(2020)\citenamefont {Zou},
  \citenamefont {Kim},\ and\ \citenamefont {Tserkovnyak}}]{zou2020tuning}%
  \BibitemOpen
  \bibfield  {author} {\bibinfo {author} {\bibfnamefont {J.}~\bibnamefont
  {Zou}}, \bibinfo {author} {\bibfnamefont {S.~K.}\ \bibnamefont {Kim}},\ and\
  \bibinfo {author} {\bibfnamefont {Y.}~\bibnamefont {Tserkovnyak}},\
  }\href@noop {} {\bibfield  {journal} {\bibinfo  {journal} {Physical Review
  B}\ }\textbf {\bibinfo {volume} {101}},\ \bibinfo {pages} {014416} (\bibinfo
  {year} {2020})}\BibitemShut {NoStop}%
\bibitem [{\citenamefont {Erlandsen}\ \emph {et~al.}(2019)\citenamefont
  {Erlandsen}, \citenamefont {Kamra}, \citenamefont {Brataas},\ and\
  \citenamefont {Sudb{\o}}}]{erlandsen2019enhancement}%
  \BibitemOpen
  \bibfield  {author} {\bibinfo {author} {\bibfnamefont {E.}~\bibnamefont
  {Erlandsen}}, \bibinfo {author} {\bibfnamefont {A.}~\bibnamefont {Kamra}},
  \bibinfo {author} {\bibfnamefont {A.}~\bibnamefont {Brataas}},\ and\ \bibinfo
  {author} {\bibfnamefont {A.}~\bibnamefont {Sudb{\o}}},\ }\href@noop {}
  {\bibfield  {journal} {\bibinfo  {journal} {Physical Review B}\ }\textbf
  {\bibinfo {volume} {100}},\ \bibinfo {pages} {100503} (\bibinfo {year}
  {2019})}\BibitemShut {NoStop}%
\bibitem [{\citenamefont {Kamra}\ \emph {et~al.}(2020)\citenamefont {Kamra},
  \citenamefont {Belzig},\ and\ \citenamefont {Brataas}}]{kamra2020magnon}%
  \BibitemOpen
  \bibfield  {author} {\bibinfo {author} {\bibfnamefont {A.}~\bibnamefont
  {Kamra}}, \bibinfo {author} {\bibfnamefont {W.}~\bibnamefont {Belzig}},\ and\
  \bibinfo {author} {\bibfnamefont {A.}~\bibnamefont {Brataas}},\ }\href@noop
  {} {\bibfield  {journal} {\bibinfo  {journal} {Applied Physics Letters}\
  }\textbf {\bibinfo {volume} {117}},\ \bibinfo {pages} {090501} (\bibinfo
  {year} {2020})}\BibitemShut {NoStop}%
\bibitem [{\citenamefont {Li}\ \emph {et~al.}(2019)\citenamefont {Li},
  \citenamefont {Zhu},\ and\ \citenamefont {Agarwal}}]{li2019squeezed}%
  \BibitemOpen
  \bibfield  {author} {\bibinfo {author} {\bibfnamefont {J.}~\bibnamefont
  {Li}}, \bibinfo {author} {\bibfnamefont {S.-Y.}\ \bibnamefont {Zhu}},\ and\
  \bibinfo {author} {\bibfnamefont {G.}~\bibnamefont {Agarwal}},\ }\href@noop
  {} {\bibfield  {journal} {\bibinfo  {journal} {Physical Review A}\ }\textbf
  {\bibinfo {volume} {99}},\ \bibinfo {pages} {021801} (\bibinfo {year}
  {2019})}\BibitemShut {NoStop}%
\bibitem [{\citenamefont {Yang}\ \emph {et~al.}(2021)\citenamefont {Yang},
  \citenamefont {Jin}, \citenamefont {Jin}, \citenamefont {Liu}, \citenamefont
  {Liu},\ and\ \citenamefont {Yang}}]{yang2021bistability}%
  \BibitemOpen
  \bibfield  {author} {\bibinfo {author} {\bibfnamefont {Z.-B.}\ \bibnamefont
  {Yang}}, \bibinfo {author} {\bibfnamefont {H.}~\bibnamefont {Jin}}, \bibinfo
  {author} {\bibfnamefont {J.-W.}\ \bibnamefont {Jin}}, \bibinfo {author}
  {\bibfnamefont {J.-Y.}\ \bibnamefont {Liu}}, \bibinfo {author} {\bibfnamefont
  {H.-Y.}\ \bibnamefont {Liu}},\ and\ \bibinfo {author} {\bibfnamefont {R.-C.}\
  \bibnamefont {Yang}},\ }\href@noop {} {\bibfield  {journal} {\bibinfo
  {journal} {Physical Review Research}\ }\textbf {\bibinfo {volume} {3}},\
  \bibinfo {pages} {023126} (\bibinfo {year} {2021})}\BibitemShut {NoStop}%
\bibitem [{\citenamefont {Zhang}\ \emph {et~al.}(2021)\citenamefont {Zhang},
  \citenamefont {Wang}, \citenamefont {Bai}, \citenamefont {Wang},
  \citenamefont {Zhang},\ and\ \citenamefont {Wang}}]{zhang2021generation}%
  \BibitemOpen
  \bibfield  {author} {\bibinfo {author} {\bibfnamefont {W.}~\bibnamefont
  {Zhang}}, \bibinfo {author} {\bibfnamefont {D.-Y.}\ \bibnamefont {Wang}},
  \bibinfo {author} {\bibfnamefont {C.-H.}\ \bibnamefont {Bai}}, \bibinfo
  {author} {\bibfnamefont {T.}~\bibnamefont {Wang}}, \bibinfo {author}
  {\bibfnamefont {S.}~\bibnamefont {Zhang}},\ and\ \bibinfo {author}
  {\bibfnamefont {H.-F.}\ \bibnamefont {Wang}},\ }\href@noop {} {\bibfield
  {journal} {\bibinfo  {journal} {Optics Express}\ }\textbf {\bibinfo {volume}
  {29}},\ \bibinfo {pages} {11773} (\bibinfo {year} {2021})}\BibitemShut
  {NoStop}%
\bibitem [{\citenamefont {Hirosawa}\ \emph {et~al.}(2022)\citenamefont
  {Hirosawa}, \citenamefont {Klinovaja}, \citenamefont {Loss},\ and\
  \citenamefont {Díaz}}]{Hirosawa2022}%
  \BibitemOpen
  \bibfield  {author} {\bibinfo {author} {\bibfnamefont {T.}~\bibnamefont
  {Hirosawa}}, \bibinfo {author} {\bibfnamefont {J.}~\bibnamefont {Klinovaja}},
  \bibinfo {author} {\bibfnamefont {D.}~\bibnamefont {Loss}},\ and\ \bibinfo
  {author} {\bibfnamefont {S.~A.}\ \bibnamefont {Díaz}},\ }\href@noop {}
  {\bibfield  {journal} {\bibinfo  {journal} {Physical Review Letters}\
  }\textbf {\bibinfo {volume} {128}},\ \bibinfo {pages} {037201} (\bibinfo
  {year} {2022})}\BibitemShut {NoStop}%
\bibitem [{\citenamefont {Owerre}(2017)}]{owerre2017floquet}%
  \BibitemOpen
  \bibfield  {author} {\bibinfo {author} {\bibfnamefont {S.}~\bibnamefont
  {Owerre}},\ }\href@noop {} {\bibfield  {journal} {\bibinfo  {journal}
  {Journal of Physics Communications}\ }\textbf {\bibinfo {volume} {1}},\
  \bibinfo {pages} {021002} (\bibinfo {year} {2017})}\BibitemShut {NoStop}%
\bibitem [{\citenamefont {Kim}\ \emph {et~al.}(2016)\citenamefont {Kim},
  \citenamefont {Ochoa}, \citenamefont {Zarzuela},\ and\ \citenamefont
  {Tserkovnyak}}]{kim2016realization}%
  \BibitemOpen
  \bibfield  {author} {\bibinfo {author} {\bibfnamefont {S.~K.}\ \bibnamefont
  {Kim}}, \bibinfo {author} {\bibfnamefont {H.}~\bibnamefont {Ochoa}}, \bibinfo
  {author} {\bibfnamefont {R.}~\bibnamefont {Zarzuela}},\ and\ \bibinfo
  {author} {\bibfnamefont {Y.}~\bibnamefont {Tserkovnyak}},\ }\href@noop {}
  {\bibfield  {journal} {\bibinfo  {journal} {Physical review letters}\
  }\textbf {\bibinfo {volume} {117}},\ \bibinfo {pages} {227201} (\bibinfo
  {year} {2016})}\BibitemShut {NoStop}%
\bibitem [{\citenamefont {Aharonov}\ \emph {et~al.}(1988)\citenamefont
  {Aharonov}, \citenamefont {Pearle},\ and\ \citenamefont
  {Vaidman}}]{aharonov1988comment}%
  \BibitemOpen
  \bibfield  {author} {\bibinfo {author} {\bibfnamefont {Y.}~\bibnamefont
  {Aharonov}}, \bibinfo {author} {\bibfnamefont {P.}~\bibnamefont {Pearle}},\
  and\ \bibinfo {author} {\bibfnamefont {L.}~\bibnamefont {Vaidman}},\
  }\href@noop {} {\bibfield  {journal} {\bibinfo  {journal} {Physical Review
  A}\ }\textbf {\bibinfo {volume} {37}},\ \bibinfo {pages} {4052} (\bibinfo
  {year} {1988})}\BibitemShut {NoStop}%
\bibitem [{\citenamefont {Meier}\ and\ \citenamefont
  {Loss}(2003)}]{meier2003magnetization}%
  \BibitemOpen
  \bibfield  {author} {\bibinfo {author} {\bibfnamefont {F.}~\bibnamefont
  {Meier}}\ and\ \bibinfo {author} {\bibfnamefont {D.}~\bibnamefont {Loss}},\
  }\href@noop {} {\bibfield  {journal} {\bibinfo  {journal} {Physical review
  letters}\ }\textbf {\bibinfo {volume} {90}},\ \bibinfo {pages} {167204}
  (\bibinfo {year} {2003})}\BibitemShut {NoStop}%
\bibitem [{\citenamefont {Elyasi}\ \emph {et~al.}(2019)\citenamefont {Elyasi},
  \citenamefont {Sato},\ and\ \citenamefont {Bauer}}]{elyasi2019topologically}%
  \BibitemOpen
  \bibfield  {author} {\bibinfo {author} {\bibfnamefont {M.}~\bibnamefont
  {Elyasi}}, \bibinfo {author} {\bibfnamefont {K.}~\bibnamefont {Sato}},\ and\
  \bibinfo {author} {\bibfnamefont {G.~E.}\ \bibnamefont {Bauer}},\ }\href@noop
  {} {\bibfield  {journal} {\bibinfo  {journal} {Physical Review B}\ }\textbf
  {\bibinfo {volume} {99}},\ \bibinfo {pages} {134402} (\bibinfo {year}
  {2019})}\BibitemShut {NoStop}%
\bibitem [{\citenamefont {Kar}\ and\ \citenamefont
  {Basu}(2018)}]{kar2018photoinduced}%
  \BibitemOpen
  \bibfield  {author} {\bibinfo {author} {\bibfnamefont {S.}~\bibnamefont
  {Kar}}\ and\ \bibinfo {author} {\bibfnamefont {B.}~\bibnamefont {Basu}},\
  }\href@noop {} {\bibfield  {journal} {\bibinfo  {journal} {Physical Review
  B}\ }\textbf {\bibinfo {volume} {98}},\ \bibinfo {pages} {245119} (\bibinfo
  {year} {2018})}\BibitemShut {NoStop}%
\bibitem [{\citenamefont {Owerre}(2019{\natexlab{a}})}]{owerre2019photo}%
  \BibitemOpen
  \bibfield  {author} {\bibinfo {author} {\bibfnamefont {S.}~\bibnamefont
  {Owerre}},\ }\href@noop {} {\bibfield  {journal} {\bibinfo  {journal} {Annals
  of Physics}\ }\textbf {\bibinfo {volume} {406}},\ \bibinfo {pages} {14}
  (\bibinfo {year} {2019}{\natexlab{a}})}\BibitemShut {NoStop}%
\bibitem [{\citenamefont {Owerre}\ \emph {et~al.}(2019)\citenamefont {Owerre},
  \citenamefont {Mellado},\ and\ \citenamefont
  {Baskaran}}]{owerre2019photoinduced}%
  \BibitemOpen
  \bibfield  {author} {\bibinfo {author} {\bibfnamefont {S.}~\bibnamefont
  {Owerre}}, \bibinfo {author} {\bibfnamefont {P.}~\bibnamefont {Mellado}},\
  and\ \bibinfo {author} {\bibfnamefont {G.}~\bibnamefont {Baskaran}},\
  }\href@noop {} {\bibfield  {journal} {\bibinfo  {journal} {EPL (Europhysics
  Letters)}\ }\textbf {\bibinfo {volume} {126}},\ \bibinfo {pages} {27002}
  (\bibinfo {year} {2019})}\BibitemShut {NoStop}%
\bibitem [{\citenamefont {Owerre}(2019{\natexlab{b}})}]{owerre2019magnonic}%
  \BibitemOpen
  \bibfield  {author} {\bibinfo {author} {\bibfnamefont {S.}~\bibnamefont
  {Owerre}},\ }\href@noop {} {\bibfield  {journal} {\bibinfo  {journal}
  {Scientific Reports}\ }\textbf {\bibinfo {volume} {9}},\ \bibinfo {pages} {1}
  (\bibinfo {year} {2019}{\natexlab{b}})}\BibitemShut {NoStop}%
\bibitem [{\citenamefont {Proskurin}\ \emph {et~al.}(2019)\citenamefont
  {Proskurin}, \citenamefont {Mac{\^e}do},\ and\ \citenamefont
  {Stamps}}]{proskurin2019microscopic}%
  \BibitemOpen
  \bibfield  {author} {\bibinfo {author} {\bibfnamefont {I.}~\bibnamefont
  {Proskurin}}, \bibinfo {author} {\bibfnamefont {R.}~\bibnamefont
  {Mac{\^e}do}},\ and\ \bibinfo {author} {\bibfnamefont {R.~L.}\ \bibnamefont
  {Stamps}},\ }\href@noop {} {\bibfield  {journal} {\bibinfo  {journal} {New
  Journal of Physics}\ }\textbf {\bibinfo {volume} {21}},\ \bibinfo {pages}
  {095003} (\bibinfo {year} {2019})}\BibitemShut {NoStop}%
\bibitem [{\citenamefont {Vinas~Bostr{\"o}m}\ \emph {et~al.}(2020)\citenamefont
  {Vinas~Bostr{\"o}m}, \citenamefont {Claassen}, \citenamefont {McIver},
  \citenamefont {Jotzu}, \citenamefont {Rubio},\ and\ \citenamefont
  {Sentef}}]{vinas2020light}%
  \BibitemOpen
  \bibfield  {author} {\bibinfo {author} {\bibfnamefont {E.}~\bibnamefont
  {Vinas~Bostr{\"o}m}}, \bibinfo {author} {\bibfnamefont {M.}~\bibnamefont
  {Claassen}}, \bibinfo {author} {\bibfnamefont {J.}~\bibnamefont {McIver}},
  \bibinfo {author} {\bibfnamefont {G.}~\bibnamefont {Jotzu}}, \bibinfo
  {author} {\bibfnamefont {A.}~\bibnamefont {Rubio}},\ and\ \bibinfo {author}
  {\bibfnamefont {M.}~\bibnamefont {Sentef}},\ }\href@noop {} {\bibfield
  {journal} {\bibinfo  {journal} {SciPost Physics}\ }\textbf {\bibinfo {volume}
  {9}},\ \bibinfo {pages} {061} (\bibinfo {year} {2020})}\BibitemShut {NoStop}%
\bibitem [{\citenamefont {Sentef}\ \emph {et~al.}(2015)\citenamefont {Sentef},
  \citenamefont {Claassen}, \citenamefont {Kemper}, \citenamefont {Moritz},
  \citenamefont {Oka}, \citenamefont {Freericks},\ and\ \citenamefont
  {Devereaux}}]{sentef2015theory}%
  \BibitemOpen
  \bibfield  {author} {\bibinfo {author} {\bibfnamefont {M.}~\bibnamefont
  {Sentef}}, \bibinfo {author} {\bibfnamefont {M.}~\bibnamefont {Claassen}},
  \bibinfo {author} {\bibfnamefont {A.}~\bibnamefont {Kemper}}, \bibinfo
  {author} {\bibfnamefont {B.}~\bibnamefont {Moritz}}, \bibinfo {author}
  {\bibfnamefont {T.}~\bibnamefont {Oka}}, \bibinfo {author} {\bibfnamefont
  {J.}~\bibnamefont {Freericks}},\ and\ \bibinfo {author} {\bibfnamefont
  {T.}~\bibnamefont {Devereaux}},\ }\href@noop {} {\bibfield  {journal}
  {\bibinfo  {journal} {Nature communications}\ }\textbf {\bibinfo {volume}
  {6}},\ \bibinfo {pages} {1} (\bibinfo {year} {2015})}\BibitemShut {NoStop}%
\bibitem [{\citenamefont {Rechtsman}\ \emph {et~al.}(2013)\citenamefont
  {Rechtsman}, \citenamefont {Zeuner}, \citenamefont {Plotnik}, \citenamefont
  {Lumer}, \citenamefont {Podolsky}, \citenamefont {Dreisow}, \citenamefont
  {Nolte}, \citenamefont {Segev},\ and\ \citenamefont
  {Szameit}}]{rechtsman2013photonic}%
  \BibitemOpen
  \bibfield  {author} {\bibinfo {author} {\bibfnamefont {M.~C.}\ \bibnamefont
  {Rechtsman}}, \bibinfo {author} {\bibfnamefont {J.~M.}\ \bibnamefont
  {Zeuner}}, \bibinfo {author} {\bibfnamefont {Y.}~\bibnamefont {Plotnik}},
  \bibinfo {author} {\bibfnamefont {Y.}~\bibnamefont {Lumer}}, \bibinfo
  {author} {\bibfnamefont {D.}~\bibnamefont {Podolsky}}, \bibinfo {author}
  {\bibfnamefont {F.}~\bibnamefont {Dreisow}}, \bibinfo {author} {\bibfnamefont
  {S.}~\bibnamefont {Nolte}}, \bibinfo {author} {\bibfnamefont
  {M.}~\bibnamefont {Segev}},\ and\ \bibinfo {author} {\bibfnamefont
  {A.}~\bibnamefont {Szameit}},\ }\href@noop {} {\bibfield  {journal} {\bibinfo
   {journal} {Nature}\ }\textbf {\bibinfo {volume} {496}},\ \bibinfo {pages}
  {196} (\bibinfo {year} {2013})}\BibitemShut {NoStop}%
\bibitem [{\citenamefont {Wang}\ \emph {et~al.}(2013)\citenamefont {Wang},
  \citenamefont {Steinberg}, \citenamefont {Jarillo-Herrero},\ and\
  \citenamefont {Gedik}}]{wang2013observation}%
  \BibitemOpen
  \bibfield  {author} {\bibinfo {author} {\bibfnamefont {Y.}~\bibnamefont
  {Wang}}, \bibinfo {author} {\bibfnamefont {H.}~\bibnamefont {Steinberg}},
  \bibinfo {author} {\bibfnamefont {P.}~\bibnamefont {Jarillo-Herrero}},\ and\
  \bibinfo {author} {\bibfnamefont {N.}~\bibnamefont {Gedik}},\ }\href@noop {}
  {\bibfield  {journal} {\bibinfo  {journal} {Science}\ }\textbf {\bibinfo
  {volume} {342}},\ \bibinfo {pages} {453} (\bibinfo {year}
  {2013})}\BibitemShut {NoStop}%
\bibitem [{\citenamefont {Lindner}\ \emph {et~al.}(2011)\citenamefont
  {Lindner}, \citenamefont {Refael},\ and\ \citenamefont
  {Galitski}}]{lindner2011floquet}%
  \BibitemOpen
  \bibfield  {author} {\bibinfo {author} {\bibfnamefont {N.~H.}\ \bibnamefont
  {Lindner}}, \bibinfo {author} {\bibfnamefont {G.}~\bibnamefont {Refael}},\
  and\ \bibinfo {author} {\bibfnamefont {V.}~\bibnamefont {Galitski}},\
  }\href@noop {} {\bibfield  {journal} {\bibinfo  {journal} {Nature Physics}\
  }\textbf {\bibinfo {volume} {7}},\ \bibinfo {pages} {490} (\bibinfo {year}
  {2011})}\BibitemShut {NoStop}%
\bibitem [{\citenamefont {Pires}(2021)}]{pires2021theoretical}%
  \BibitemOpen
  \bibfield  {author} {\bibinfo {author} {\bibfnamefont {A.~S.~T.}\
  \bibnamefont {Pires}},\ }\href@noop {} {\emph {\bibinfo {title} {Theoretical
  Tools for Spin Models in Magnetic Systems}}}\ (\bibinfo  {publisher} {IOP
  Publishing},\ \bibinfo {year} {2021})\BibitemShut {NoStop}%
\bibitem [{\citenamefont {Kamra}\ \emph {et~al.}(2017)\citenamefont {Kamra},
  \citenamefont {Agrawal},\ and\ \citenamefont {Belzig}}]{kamra2017noninteger}%
  \BibitemOpen
  \bibfield  {author} {\bibinfo {author} {\bibfnamefont {A.}~\bibnamefont
  {Kamra}}, \bibinfo {author} {\bibfnamefont {U.}~\bibnamefont {Agrawal}},\
  and\ \bibinfo {author} {\bibfnamefont {W.}~\bibnamefont {Belzig}},\
  }\href@noop {} {\bibfield  {journal} {\bibinfo  {journal} {Physical Review
  B}\ }\textbf {\bibinfo {volume} {96}},\ \bibinfo {pages} {020411} (\bibinfo
  {year} {2017})}\BibitemShut {NoStop}%
\bibitem [{\citenamefont {Harder}(2018)}]{harder2018cavity}%
  \BibitemOpen
  \bibfield  {author} {\bibinfo {author} {\bibfnamefont {M.}~\bibnamefont
  {Harder}},\ }\emph {\bibinfo {title} {Cavity spintronics: foundations and
  applications of spin-photon hybridization}},\ \href@noop {} {Ph.D. thesis},\
  \bibinfo  {school} {Faculty of Graduate Studies of The University of
  Manitoba} (\bibinfo {year} {2018})\BibitemShut {NoStop}%
\bibitem [{\citenamefont {Mahan}(2000)}]{MahanManyParticle}%
  \BibitemOpen
  \bibfield  {author} {\bibinfo {author} {\bibfnamefont {G.~D.}\ \bibnamefont
  {Mahan}},\ }\href@noop {} {\emph {\bibinfo {title} {Many-Particle
  Physics}}},\ \bibinfo {edition} {3rd}\ ed.\ (\bibinfo  {publisher} {Springer
  Science+Business Media},\ \bibinfo {address} {New York},\ \bibinfo {year}
  {2000})\BibitemShut {NoStop}%
\bibitem [{\citenamefont {Gong}\ \emph {et~al.}(2018)\citenamefont {Gong},
  \citenamefont {Gong}, \citenamefont {Sun}, \citenamefont {Tong},
  \citenamefont {Duan}, \citenamefont {Chu},\ and\ \citenamefont
  {Zhang}}]{gong2018electrically}%
  \BibitemOpen
  \bibfield  {author} {\bibinfo {author} {\bibfnamefont {S.-J.}\ \bibnamefont
  {Gong}}, \bibinfo {author} {\bibfnamefont {C.}~\bibnamefont {Gong}}, \bibinfo
  {author} {\bibfnamefont {Y.-Y.}\ \bibnamefont {Sun}}, \bibinfo {author}
  {\bibfnamefont {W.-Y.}\ \bibnamefont {Tong}}, \bibinfo {author}
  {\bibfnamefont {C.-G.}\ \bibnamefont {Duan}}, \bibinfo {author}
  {\bibfnamefont {J.-H.}\ \bibnamefont {Chu}},\ and\ \bibinfo {author}
  {\bibfnamefont {X.}~\bibnamefont {Zhang}},\ }\href@noop {} {\bibfield
  {journal} {\bibinfo  {journal} {Proceedings of the National Academy of
  Sciences}\ }\textbf {\bibinfo {volume} {115}},\ \bibinfo {pages} {8511}
  (\bibinfo {year} {2018})}\BibitemShut {NoStop}%
\bibitem [{\citenamefont {Yoon}\ \emph {et~al.}(2021)\citenamefont {Yoon},
  \citenamefont {Kim}, \citenamefont {Choi}, \citenamefont {Sung},
  \citenamefont {Lee}, \citenamefont {Lee},\ and\ \citenamefont
  {Nam}}]{yoon2021realization}%
  \BibitemOpen
  \bibfield  {author} {\bibinfo {author} {\bibfnamefont {J.~W.}\ \bibnamefont
  {Yoon}}, \bibinfo {author} {\bibfnamefont {Y.~G.}\ \bibnamefont {Kim}},
  \bibinfo {author} {\bibfnamefont {I.~W.}\ \bibnamefont {Choi}}, \bibinfo
  {author} {\bibfnamefont {J.~H.}\ \bibnamefont {Sung}}, \bibinfo {author}
  {\bibfnamefont {H.~W.}\ \bibnamefont {Lee}}, \bibinfo {author} {\bibfnamefont
  {S.~K.}\ \bibnamefont {Lee}},\ and\ \bibinfo {author} {\bibfnamefont {C.~H.}\
  \bibnamefont {Nam}},\ }\href@noop {} {\bibfield  {journal} {\bibinfo
  {journal} {Optica}\ }\textbf {\bibinfo {volume} {8}},\ \bibinfo {pages} {630}
  (\bibinfo {year} {2021})}\BibitemShut {NoStop}%
\bibitem [{\citenamefont {Noginov}\ \emph {et~al.}(2009)\citenamefont
  {Noginov}, \citenamefont {Zhu}, \citenamefont {Belgrave}, \citenamefont
  {Bakker}, \citenamefont {Shalaev}, \citenamefont {Narimanov}, \citenamefont
  {Stout}, \citenamefont {Herz}, \citenamefont {Suteewong},\ and\ \citenamefont
  {Wiesner}}]{noginov2009demonstration}%
  \BibitemOpen
  \bibfield  {author} {\bibinfo {author} {\bibfnamefont {M.}~\bibnamefont
  {Noginov}}, \bibinfo {author} {\bibfnamefont {G.}~\bibnamefont {Zhu}},
  \bibinfo {author} {\bibfnamefont {A.}~\bibnamefont {Belgrave}}, \bibinfo
  {author} {\bibfnamefont {R.}~\bibnamefont {Bakker}}, \bibinfo {author}
  {\bibfnamefont {V.}~\bibnamefont {Shalaev}}, \bibinfo {author} {\bibfnamefont
  {E.}~\bibnamefont {Narimanov}}, \bibinfo {author} {\bibfnamefont
  {S.}~\bibnamefont {Stout}}, \bibinfo {author} {\bibfnamefont
  {E.}~\bibnamefont {Herz}}, \bibinfo {author} {\bibfnamefont {T.}~\bibnamefont
  {Suteewong}},\ and\ \bibinfo {author} {\bibfnamefont {U.}~\bibnamefont
  {Wiesner}},\ }\href@noop {} {\bibfield  {journal} {\bibinfo  {journal}
  {Nature}\ }\textbf {\bibinfo {volume} {460}},\ \bibinfo {pages} {1110}
  (\bibinfo {year} {2009})}\BibitemShut {NoStop}%
\bibitem [{\citenamefont {Harder}\ \emph {et~al.}(2018)\citenamefont {Harder},
  \citenamefont {Yang}, \citenamefont {Yao}, \citenamefont {Yu}, \citenamefont
  {Rao}, \citenamefont {Gui}, \citenamefont {Stamps},\ and\ \citenamefont
  {Hu}}]{harder2018level}%
  \BibitemOpen
  \bibfield  {author} {\bibinfo {author} {\bibfnamefont {M.}~\bibnamefont
  {Harder}}, \bibinfo {author} {\bibfnamefont {Y.}~\bibnamefont {Yang}},
  \bibinfo {author} {\bibfnamefont {B.}~\bibnamefont {Yao}}, \bibinfo {author}
  {\bibfnamefont {C.}~\bibnamefont {Yu}}, \bibinfo {author} {\bibfnamefont
  {J.}~\bibnamefont {Rao}}, \bibinfo {author} {\bibfnamefont {Y.}~\bibnamefont
  {Gui}}, \bibinfo {author} {\bibfnamefont {R.}~\bibnamefont {Stamps}},\ and\
  \bibinfo {author} {\bibfnamefont {C.-M.}\ \bibnamefont {Hu}},\ }\href@noop {}
  {\bibfield  {journal} {\bibinfo  {journal} {Physical review letters}\
  }\textbf {\bibinfo {volume} {121}},\ \bibinfo {pages} {137203} (\bibinfo
  {year} {2018})}\BibitemShut {NoStop}%
\bibitem [{\citenamefont {Yao}\ \emph {et~al.}(2019)\citenamefont {Yao},
  \citenamefont {Yu}, \citenamefont {Zhang}, \citenamefont {Lu}, \citenamefont
  {Gui}, \citenamefont {Hu},\ and\ \citenamefont
  {Blanter}}]{yao2019microscopic}%
  \BibitemOpen
  \bibfield  {author} {\bibinfo {author} {\bibfnamefont {B.}~\bibnamefont
  {Yao}}, \bibinfo {author} {\bibfnamefont {T.}~\bibnamefont {Yu}}, \bibinfo
  {author} {\bibfnamefont {X.}~\bibnamefont {Zhang}}, \bibinfo {author}
  {\bibfnamefont {W.}~\bibnamefont {Lu}}, \bibinfo {author} {\bibfnamefont
  {Y.}~\bibnamefont {Gui}}, \bibinfo {author} {\bibfnamefont {C.-M.}\
  \bibnamefont {Hu}},\ and\ \bibinfo {author} {\bibfnamefont {Y.~M.}\
  \bibnamefont {Blanter}},\ }\href@noop {} {\bibfield  {journal} {\bibinfo
  {journal} {Physical Review B}\ }\textbf {\bibinfo {volume} {100}},\ \bibinfo
  {pages} {214426} (\bibinfo {year} {2019})}\BibitemShut {NoStop}%
\end{thebibliography}

\end{document}